\documentclass[11pt, a4paper]{article}

\usepackage[oldstyle]{hep-font} %

\usepackage{fullpage}

\usepackage{hep-math-font} %
\usepackage{hep-math} %
\numberwithin{equation}{section}

\usepackage{hep-bibliography} %
\bibliography{bibliography}

\usepackage{hep-float} %

\usepackage{hep-text} %

\usepackage{accsupp}
\newcommand\numberstyle[1]{\tstyle\BeginAccSupp{ActualText={}}#1\protect\EndAccSupp{}\small}
\usepackage{listings} %
\usepackage{hep-reference} %

\usepackage{hep-acronym} %

\usepackage{hep-title} %

\usepackage[feynman,plot]{hep-graphic}
\graphicpath{./graphics}
\pgfplotsset{table/search path={data}}
\usetikzlibrary{positioning}
\usetikzlibrary{fit}
\usetikzlibrary{shapes}
\usetikzlibrary{patterns}
\usetikzlibrary{patterns.meta}
\usepgfplotslibrary{fillbetween}

\NewDocumentCommand{\scatterLNV}{omm}{{
\newcommand{\conj}{$\IfValueTF{#1}{#1}{#2}$}
\newcommand{\yuk}{$#2$}
\newcommand{\name}{$\LNV$}
\newcommand{\mass}{$#3$}
\begin{tikzpicture}[
baseline={([yshift=-.5ex]current bounding box.center)},
]
\begin{feynman}[
small,
node distance = 3ex and 1.5em,
]
\vertex (v1);
\vertex (i1) [above left = of v1];
\vertex (i2) [below left = of v1];
\vertex (v2) [right = of v1];
\vertex (f1) [above right = of v2];
\vertex (f2) [below right = of v2];
\diagram*{
 {%
  (i1) -- (v1) --[edge label=\name,edge label'=\mass] (v2) -- (f1),
 }, {[edges = boson]
  (i2) -- (v1),
  (f2) -- (v2),
 }
};
\node[left=.8em of v1,inner sep=1pt, anchor=east]{\yuk};
\node[right=.8em of v2,inner sep=1pt,color=black!50,anchor=west]{\conj};
\end{feynman}
\end{tikzpicture}
}}
\NewDocumentCommand{\scatterLNC}{omm}{{
\newcommand{\conj}{$\IfValueTF{#1}{#1}{#2^*}$}
\newcommand{\yuk}{$#2$}
\newcommand{\name}{$\LNC$}
\newcommand{\mass}{$#3$}
\begin{tikzpicture}[
baseline={([yshift=-.5ex]current bounding box.center)},
]
\begin{feynman}[
small,
node distance = 3ex and 1.5em,
]
\vertex (v1);
\vertex (i1) [above left = of v1];
\vertex (i2) [below left = of v1];
\vertex (v2) [right = of v1];
\vertex (f1) [above right = of v2];
\vertex (f2) [below right = of v2];
\diagram*{
 {%
  (i1) -- (v1) --[edge label=\name,edge label'=\mass] (v2) -- (f1),
 }, {[edges = boson]
  (i2) -- (v1),
  (f2) -- (v2),
 }
};
\node[left=.8em of v1,inner sep=1pt, anchor=east]{\yuk};
\node[right=.8em of v2,inner sep=1pt,color=black!50,anchor=west]{\conj};
\end{feynman}
\end{tikzpicture}
}}
\NewDocumentCommand{\coupling}{ommm}{{
\newcommand{\boson}{$\IfValueTF{#1}{#1}{W}$}
\newcommand{\heavy}{$#2$}
\newcommand{\light}{$#3$}
\newcommand{\yuk}{$#4$}
\begin{tikzpicture}[
baseline={([yshift=-.5ex]current bounding box.center)},
]
\begin{feynman}[
small,
node distance = 3ex and 1.5em,
]
\vertex (v1);
\vertex (i1) [above left = of v1];
\vertex (i2) [below left = of v1];
\vertex (f1) [right = of v1];

\diagram*{
 {%
  (i1) -- (v1) -- (i2),
 }, {[edges = boson]
  (v1) -- (f1),
 }
};
\node[left=.8em of v1,inner sep=1pt,anchor=east]{\yuk};
\node[left=.4em of i1,inner sep=1pt,anchor=east]{\heavy};
\node[left=.4em of i2,inner sep=1pt,anchor=east]{\light};
\node[right=.4em of f1,inner sep=1pt,anchor=west]{\boson};
\end{feynman}
\end{tikzpicture}
}}

\acronym{HNL}{heavy neutral lepton}
\acronym{LNC}{lepton number conservation}
\acronym{LNV}{lepton number violation}
\acronym{SM}{Standard Model}
\acronym{GUT}{grand unified theory}[grand unified theories]
\acronym{QFT}{quantum field theoretical}
\acronym{QM}{quantum mechanical}
\acronym{EW}{electroweak}
\acronym{EWSB}{electroweak symmetry breaking}
\acronym{VEV}{vacuum expectation value}
\acronym{SPSS}{symmetry protected seesaw scenario}
\acronym{LHC}{large hadron collider}
\acronym[HL-LHC]{HLLHC}{high luminosity phase of the \LHC}
\acronym[LH$e$C]{LHeC}{large hadron electron collider}
\shortacronym{FCC}*{future circular collider}
\shortacronym{ATLAS}{A Toroidal \LHC ApparatuS}
\acronym{HL}{high luminosity}
\acronym{DOF}{degree of freedom}[degrees of freedom]
\acronym{LNLS}{\emph{lepton number}-like symmetry}
\shortacronym{CMS}{compact muon solenoid}
\acronym{LO}{leading order}
\acronym{CL}{charged lepton}
\acronym{NP}{new physics}
\acronym{MC}{Monte Carlo}
\acronym{PMNS}{Pontecorvo–Maki–Nakagawa–Sakata}
\acronym{pSPSS}{phenomenological symmetry protected seesaw scenario}
\acronym[$N\mkern-1.5mu\widebar N$O]{NNO}{heavy neutrino-antineutrino oscillation}
\acronym{PDF}{probability density function}
\acronym{LLR}{logarithm of the likelihood ratio}
\shortacronym{UFO}{universal \software{FeynRules} output}
\acronym[H.c.]{HC}{Hermitian conjugate}
\acronym[$0\nu$\beta\beta]{NLDB}{neutrinoless double \beta}
\acronym{BM}{benchmark model}
\acronym{TOF}{time of flight}

\renewcommand{\vec}{\mathbf}
\NewDocumentCommand{\expno}{om}{\IfNoValueF{#1}{#1\times}10^{#2}}
\newcommand{\expnumber}[2]{\expno[#1]{#2}}
\DeclareMathOperator{\SU}{SU}
\DeclareMathOperator{\BR}{BR}
\newcommand{\ispd}{\differential{i\slashed\partial}}

\title{Simulating lepton number violation induced by heavy neutrino-antineutrino oscillations at colliders}

\author[a]{Stefan Antusch \email{stefan.antusch@unibas.ch}}
\author[b]{Jan Hajer \email{jan.hajer@tecnico.ulisboa.pt}}
\author[a]{Johannes Rosskopp \email{johannes.rosskopp@unibas.ch}}

\affiliation[a]{Departement Physik, Universität Basel, Klingelbergstrasse 82, CH-4056 Basel, Switzerland}
\affiliation[b]{Centro de Física Teórica de Partículas (CFTP), Instituto Superior Técnico (IST), Universidade de Lisboa, Av.\ Rovisco Pais 1, 1049-001 Lisboa, Portugal}

\begin{document}

\maketitle

\begin{abstract}
We study pseudo-Dirac pairs of two almost mass-degenerate sterile Majorana neutrinos which generate light neutrino masses via a low-scale seesaw mechanism.
These pseudo-Dirac heavy neutral leptons can oscillate between interaction eigenstates that couple to leptons and antileptons and thus generate oscillations between lepton number conserving and lepton number violating processes.
With the \pSPSS, we introduce a minimal framework capable of describing the dominant features of low-scale seesaws at colliders and present a \software{FeynRules} implementation usable in Monte Carlo generators.
Additionally, we extend \software{MadGraph} to simulate heavy neutrino-antineutrino oscillations and present results from such simulations.
\end{abstract}

\clearpage
\tableofcontents
\clearpage
\listoffigures
\listoftables
\lstlistoflistings
\clearpage

\section{Introduction}

The discovery of flavour oscillations between \SM neutrinos \cite{SNO:2002tuh} implies that they have nonzero masses.
One possible extension of the \SM able to generate neutrino masses consists of adding sterile neutrinos, that are singlets under all \SM symmetries, to its particle content \cite{Minkowski:1977sc,Gell-Mann:1979vob,Mohapatra:1979ia,Glashow:1979nm,Yanagida:1980xy,Schechter:1980gr,Schechter:1981cv}.
Testing whether these sterile neutrinos are indeed the \emph{missing piece} that explains the light neutrino masses is one of the key questions towards a more complete theory of elementary particles.

With sterile neutrinos added to the \SM, there are two main routes resulting in nonzero neutrino masses.
In the first case, only Dirac mass terms are present.
The Yukawa coupling term is formed by sterile neutrinos, the Higgs field and the lepton doublets, analogous to the mechanism generating charged fermion masses.
After \EWSB, this Yukawa term generates Dirac neutrino masses.
However, to yield the phenomenologically required small masses for the light neutrinos, the size of the Yukawa couplings has to be tiny $\vec y \lessapprox \order{\expno{-12}}$.

Since the sterile neutrinos are \SM singlets, it is possible to add a Majorana mass term to the \SM Lagrangian \cite{Majorana:1937vz}.
Therefore, realising light Dirac neutrinos requires a mechanism to eliminate this term and enforce exact \LNC.
As soon as some \LNV is present, the light neutrinos are of Majorana-type \cite{Schechter:1981bd}.
One way to probe these considerations are searches for \NLDB decay \cite{Furry:1939qr} since its observation would prove \LNV.
The two main routes for nonzero neutrino masses are depicted in the upper part of \cref{fig:neutrino types}, and in this work we focus on the case that the light neutrinos have Majorana masses.

As an alternative to \NLDB decays, one can also search for \LNV induced by the sterile neutrinos at colliders.
While this sounds promising at first sight, one can argue on general grounds that \LNV, observable at the \LHC, would lead to too heavy light neutrino masses, and thus it should be impossible to observe \LNV \cite{Kersten:2007vk}.
An effect that has not been taken into account in such considerations are \NNOs \cite{Antusch:2020pnn}, see also \cite{Cvetic:2015ura,Anamiati:2016uxp,Antusch:2017ebe}.
In particular, their interplay with the potential longevity of the heavy neutrinos is the main subject of this paper.
Despite the smallness of the \LNV terms in the Lagrangian, \NNOs can introduce \LNV processes at the same order as \LNC processes, depending on the lifetime and oscillation period of the sterile neutrinos.
Since the oscillations are an interference phenomenon, they are able to probe mass splittings of heavy neutrinos so small that they would otherwise be unobservable in collider studies.
Including the \NNOs correctly in collider studies is thus a crucial aspect when simulating seesaw extensions of the \SM to explain the light neutrino masses.

This paper is organised as follows: In \cref{sec:seesaw}, we describe the possible domains of seesaw models and argue that collider testable seesaw models are protected by a \LNLS resulting in pseudo-Dirac pairs of heavy neutrinos.
Afterwards, in \cref{sec:oscillations}, we give the relevant results for the description of \NNOs in the external wave packet formalism and derive the integrated effects of these oscillations.
Subsequently, we introduce the \SPSS \cite{Antusch:2015mia,Antusch:2016ejd} in \cref{sec:SPSS}, first in the symmetric limit, then extended by small symmetry violating terms, and finally as the \pSPSS with the minimal set of parameters able to describe the dominant collider effects of low-scale type I seesaws.
In \cref{sec:implementation}, we introduce the \software{FeynRules} model file of the \pSPSS and describe the necessary steps to extend \software{MadGraph} to be able to simulate \NNOs.
In \cref{sec:results}, we present selected results from a \MC study using this software implementation.
Finally, we conclude in \cref{sec:conclusion}.
Additionally, we comment in \cref{sec:HNL characterisation} on the discussion about whether it is possible to distinguish Majorana and Dirac particles from each other using their decay width.
Last but not least, the code for the \software{MadGraph} patch is presented in \cref{sec:patch}.

\resetacronym{SPSS}
\resetacronym{pSPSS}
\resetacronym{NNO}
\resetacronym{LNLS}

\section{Seesaw models} \label{sec:seesaw}

\begin{figure}
\includetikz{neutrinos}
\caption[Comparison between the different possible choices for active and sterile neutrinos]{
Comparison between the different possible choices for active and sterile neutrinos discussed in the text.
Note that for collider-accessible heavy neutrinos that generate Majorana masses for the light neutrinos via a low-scale seesaw with couplings far above the naive seesaw line, only pseudo-Dirac pairs of two nearly mass-degenerate Majorana $\DOFs$ are a viable option.
} \label{fig:neutrino types}
\end{figure}

With $i = 1, \dots, n$ sterile neutrinos $N_i$ and both Dirac and Majorana mass terms extending the \SM Lagrangian, one arrives at a general theoretical framework where the sterile neutrino Lagrangian
\footnote{
For ease of notation, the sterile neutrinos are introduced as left-chiral $\DOFs$.
Note that any left-chiral field can also be described by the charge conjugate of a right-chiral field and vice versa.
}
\begin{equation} \label{eq:generic seesaw}
\mathcal L_N =
- \sum_i y_\alpha^{(i)} \widebar{N_i^c} \widetilde H^\dagger \ell_\alpha
- \frac12 \sum_i m_M^{(i)} \widebar{N_i^c} N_i^{}
+ \HC \,,
\end{equation}
is added to the \SM Lagrangian.
Here $H$ and $\ell$ are the \SM Higgs and lepton $\SU(2)$ doublets, respectively, $\vec y^{(i)}$ is the neutrino Yukawa coupling vector $\row{y_1^{(i)},y_2^{(i)},y_3^{(i)}}^\trans$, and $m_M^{(i)}$ is the Majorana mass parameter.
\footnote{
We indicate quantities with a suppressed vectorial index by using boldface font.
}
To illustrate how small neutrino masses can emerge from this framework, we consider the minimal case of two sterile neutrinos $n=2$.
It is minimal since neutrino flavour oscillations require at least two of the light neutrino masses to be nonzero, which in turn requires at least two sterile neutrino \DOFs.
The following discussion can be generalised to larger $n$, which is necessary for models such as the inverse seesaw to be phenomenological viable.
After \EWSB, the Lagrangian can be written in the diagonal basis for the Majorana masses $m_M^{}$,
\begin{equation} \label{eq:mass Lagrangian}
\mathcal L_N =
- m_{D\alpha}^{(1)} \widebar{N_1^c} \nu_\alpha
- m_{D\alpha}^{(2)} \widebar{N_2^c} \nu_\alpha
- \frac12 m_M^{(1)} \widebar{N_1^c} N_1^{}
- \frac12 m_M^{(2)} \widebar{N_2^c} N_2^{}
+ \HC \,,
\end{equation}
where $\vec m_D^{} = \vec y v$ with the \SM Higgs \VEV $v \approx \unit[174]{GeV}$ describes the Dirac mass contribution.
When $m_M^{}$ is sufficiently larger than $\vec m_D^{}$, the light neutrino mass matrix is given by the seesaw relation, which takes the form
\begin{equation} \label{eq:seesaw}
M_\nu = \frac{\vec m_D^{(1)} \otimes \vec m_D^{(1)}}{m_M^{(1)}} + \frac{\vec m_D^{(2)} \otimes \vec m_D^{(2)}}{m_M^{(2)}} \,.
\end{equation}
With these ingredients, there are three limiting cases allowing to arrive at the observed small neutrino masses:

\begin{figure}
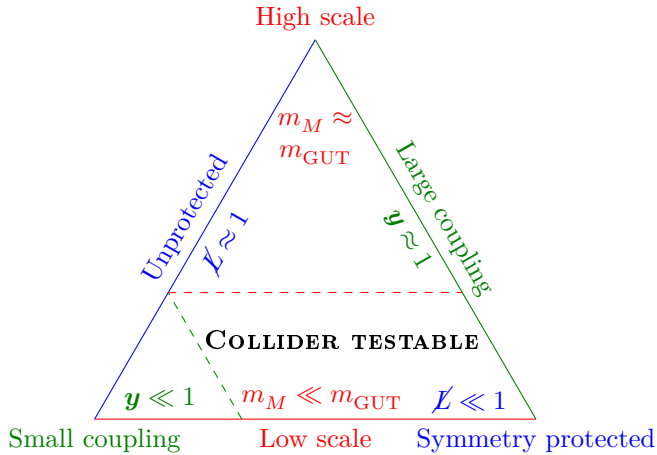

\includetikz*{triangle}
\caption[Limiting cases of seesaw models]{
Depiction of the three limiting seesaw regimes described in the text.
The phenomenologically accessible parameter space of low-scale seesaws with large enough coupling is connected to the symmetry-protected corner and indicated by dashed lines.
} \label{fig:seesaw triangle}
\end{figure}

\begin{enumdescript}[label=\roman*)]

\item{High-scale seesaw limit}

Here neutrino masses are small because the Majorana masses $m_M^{}$ of the sterile neutrinos are large.
The neutrino Yukawa couplings $\vec y$ can be large in this case --- up to $\order 1$ for $m_M^{}$ close to the scale of grand unification.
Due to the high mass scale involved, this limit of the seesaw mechanism cannot be probed directly at colliders.

\item{Small coupling seesaw limit}

It is possible to simultaneously lower the sterile neutrino mass scale $m_M^{}$ and the size of the neutrino Yukawa couplings $\vec y$ without changing the generated light neutrino masses $M_\nu$ in \cref{eq:seesaw}.
Envisioned future collider experiments such as the \FCC-$ee$ can probe this limit for some choices of parameters, but for most cases the couplings are too small for direct tests.

\item{Symmetry-protected seesaw limit}

The third possibility emerges when the two terms in \cref{eq:seesaw} almost cancel.
Then $\vec y$ can be large enough, and simultaneously $m_M^{}$ can be small enough, such that the heavy neutrinos are within reach of collider experiments.
This cancellation can be protected by a \LNLS that generalises the lepton number $L$ of the \SM and ensures that $M_\nu$ is equal to zero in the symmetry conserving limit.
\footnote{
One can achieve this cancellation also by tuning of parameters without relying on a symmetry.
However, this mechanism of cancellation is unstable under radiative corrections, \cf \cite{Kersten:2007vk}.
We will not consider this possibility further here, only noting that in this case, it is expected that the \LNV effects would be generically unsuppressed.
}

\end{enumdescript}
The various manifestations of the \emph{seesaw mechanism} are schematically depicted in \cref{fig:seesaw triangle}.
The three corners represent the three limiting cases discussed above.
However, also the cases in between are viable options.
For the Majorana mass scales $m_M^{}$, there is a maximal value above which it is no longer possible to test this sterile neutrino directly at a given collider experiment.
This means the observable seesaw models are \emph{low-scale models} opposite to the high-scale seesaw corner.
Among the low-scale seesaw models, the small coupling limit is also not testable at \eg the \LHC or even the \HLLHC.
The potentially \emph{testable} region of the (type I) seesaw mechanism is thus the area inside the dashed lines in \cref{fig:seesaw triangle}; it is bound to have a certain degree of symmetry protection if one wants to avoid tuning of the parameters.

Although the model of two sterile neutrinos with exact \LNLS serves as a valuable starting point for the construction of viable seesaw models with small symmetry breaking, it is itself not able to generate neutrino masses due to the unbroken symmetry.
With the symmetry intact, the two Majorana \DOFs are mass degenerate and combine precisely to form a single Dirac particle.
However, a small amount of symmetry breaking not only generates small neutrino masses but at the same time causes a small mass splitting between the two Majorana \DOFs.
Such pseudo-Dirac heavy neutrinos can exhibit \NNO potentially detectable at collider experiments.
For large symmetry breaking, this feature vanishes as the mass splitting becomes too big and decoherence sets in so that the two sterile neutrinos appear as two separate Majorana particles.
At the same time, the generated \SM neutrino masses become too large if the model has no additional mechanism to prevent this from happening.
We depict these considerations in the lower part of \cref{fig:neutrino types}.

Although realistic low-scale seesaw models predict pseudo-Dirac heavy neutrinos, the majority of searches for \HNLs have been performed in either the pure Dirac or single Majorana scenario.
In an effort to distinguish between these two models, it has sometimes been argued that one can discriminate Majorana from Dirac \HNLs using their decay width.
In \cref{sec:HNL characterisation}, we describe in detail why this is not the case.
The main insight is that the factor of two appearing when comparing the decay widths counts the number of Majorana \DOFs forming the observed \HNL.
However, two Majorana particles can only be described as a Dirac particle when their Yukawa couplings have a relative phase of $-i$, which cannot be determined using the decay width, \cf \cref{sec:HNL characterisation}.

\section{\sentence\NNOslong} \label{sec:oscillations}

\resetacronym{NNO}

\begin{figure}
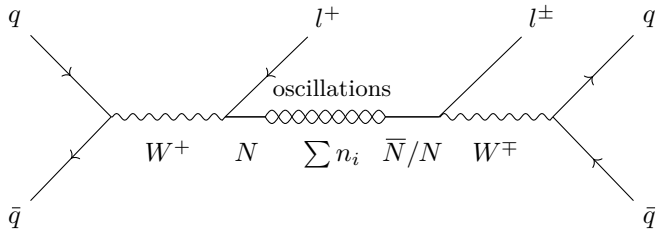

\includetikz*{oscillation-feynman}
\caption[Feynman diagram visualising the oscillations]{
A neutrino interaction eigenstate is produced together with an antilepton.
After some time, the superposition of mass eigenstates $n_i$ has oscillated into a heavy neutrino $N$ or antineutrino $\widebar N$ interaction eigenstate that decays into an antilepton $l^+$ or lepton $l^-$, such that the total process is \LNV or \LNC.
} \label{fig:oscillation feynman}
\end{figure}

\begin{figure}
\begin{panels}[t]{3}
\includetikz{oscillation-short}
\caption{Short oscillation period.} \label{fig:short}
\panel
\includetikz{oscillation}
\caption{Intermediate oscillation period.} \label{fig:Intermediate}
\panel
\includetikz{oscillation-long}
\caption{Long oscillation period.} \label{fig:long}
\end{panels}
\caption[Depiction of oscillation periods]{
Oscillations of the probability $P$ that the process shown in \cref{fig:oscillation feynman} is \LNC or \LNV as a function of the proper time $\tau$.
The oscillation period depends on the mass splitting $\Delta m$, the overall decay depends on the decay width $\Gamma$, and the decoherence depends on the damping parameter $\lambda$, all of which are defined in \cref{sec:lo oscillations}.
Panel \subref{fig:short} shows a short oscillation period with $\Delta m/\Gamma = 100$, panel \subref{fig:Intermediate} shows an intermediate oscillation period with $\Delta m/\Gamma = 10$, and panel \subref{fig:long} shows a long oscillation period with $\Delta m/\Gamma = 1$.
All oscillations are shown for $\Gamma \approx \frac{\unit{meV}}{25}$ and $\lambda = \nicefrac15$ and $R_{ll}$ is defined in \cref{sec:integrated effect}.
} \label{fig:oscillations}
\end{figure}

Generically, neutral particles can oscillate into their antiparticles \cite{Gell-Mann:1955ipe,Pontecorvo:1957cp,Pontecorvo:1967fh} as known \eg from meson oscillations \cite{Lande:1956pf,ARGUS:1987xtv,CDF:2006imy,LHCb:2012zll}.
The distinction between particle and antiparticle in the case of sterile neutrinos can be made by distinguishing a neutrino interaction eigenstate by the charge of the associated lepton it is produced with.
If it is produced together with a lepton $l^-$, it is an antineutrino and if it is produced together with an anti\-lepton~$l^+$, it is a neutrino.
The corresponding mass eigenstates $n_i$ interfere when propagating, which results in an oscillation between different interaction eigenstates as a function of travelling distance.
Heavy neutrinos~$N$ and antineutrinos $\widebar N$ are defined as the projection of neutrino and antineutrino states onto the heavy mass eigenstates.
This mechanism leads to \NNOs.
In the collider testable region of \cref{fig:seesaw triangle}, the mass splitting can be small enough for the \NNOs to be observable and of macroscopic length.
If \NNOs are present, one expects processes such as the one shown in \cref{fig:oscillation feynman} to occur, leading to patterns as depicted in \cref{fig:oscillations}.
Furthermore, in cases in which the parameters do not allow for macroscopic oscillations to be resolvable, an integrated effect could still be measured.

\subsection{External wave packets} \label{sec:external wave packets}

In order to predict the oscillatory behaviour of new particles, a holistic \QFT framework is necessary that, in contrast to the simplified \QM framework, is able to not only capture potential oscillations but can also predict the potential damping of these oscillations due to the loss of coherence of the mass eigenstate superposition.
Compared to the plane wave description of particle oscillations, the \QFT external wave packet approach discussed in \cite{Beuthe:2001rc} and adapted to the case of \NNOs in \cite{Antusch:2020pnn} allows the derivation of the phenomenon free of contradictions.
This is due to the fact that the inherent uncertainty of wave packets in momentum and position space allows the simultaneous production of several on-shell mass eigenstates that can subsequently interfere to produce oscillations.
Furthermore, the approximate localisation of wave packets is necessary to introduce the notion of a travelled distance and time, which is not possible using infinitely extended plane waves.
Additionally, several effects potentially leading to the decoherence of the mass eigenstate superposition are included and discussed under the name of \emph{observability conditions} \cite{Antusch:2023nqd}.
\emph{External} refers to the fact that only the external particles are explicitly assumed to be wave packets, whereas in an intermediate wave packed approach, the intermediate particles, such as in this case the heavy neutrinos, are directly described by wave packets.
However, even in the external wave packet approach, it is possible to interpret the intermediate particles as wave packets where some clarifications about the causal nature of this interpretation can be found in \cite{Beuthe:2001rc,Giunti:2002xg}.

For the external wave packet models, the assumptions about the shape and the widths of the involved wave packets are a major source of uncertainty when deducing results.
Also, the formalism discussed in \cite{Beuthe:2001rc,Antusch:2020pnn} was mainly developed having long-lived quasi mass-degenerate particles in mind.
In particular, the theorem of Jacob-Sachs \cite{Jacob:1961zz} is only valid above a certain time threshold, below which additional corrections have to be taken into account.
In \cite{Antusch:2020pnn,Antusch:2023nqd}, the regions, as well as the effects of the above-mentioned corrections, are discussed in detail, and it is shown that in phenomenological studies of nearly mass-degenerate, quasi long-lived heavy neutrinos, the effects of the corrections can be neglected for a range of external wave packet widths spanning orders of magnitude.

\subsection{Oscillations at \LOlong} \label{sec:lo oscillations}

In the case of slightly broken \LNLS governing the physics of pseudo-Dirac heavy neutrinos, the complete \LO contributions to the \NNO can be described using only two parameters in addition to the seesaw model in the \LNC limit \cite{Antusch:2020pnn,Antusch:2023nqd}.
Hence instead of working with the whole family of realistic seesaw models and their parameters, one can instead introduce, in addition to the heavy neutrino mass $m$ and the active-sterile mixing parameter $\vec\theta$ defined in \cref{eq:active-sterile mixing}, just the mass splitting of the heavy neutrinos $\Delta m$ and the damping parameter $\lambda$.
As an effective parameter, $\Delta m$ captures the effects induced by the \LNLS breaking parameters, and $\lambda$ entails the decoherence effects appearing in certain parameter regions of the \QFT description of neutral particle oscillations.
To \LO, the formulae for the oscillation probabilities as a function of the proper time $\tau$ are
\begin{equation}
P^{\nicefrac{\LNC}{\LNV}}_\text{osc}(\tau) = \frac{1 \pm \cos\left(\Delta m \tau \right) \exp(-\lambda)}{2}\,,
\end{equation}
hence the oscillation period is given by $\tau_\text{osc} = 2 \pi / \Delta m$.
The overall decay of the heavy neutrino is given by the usual probability density for the decay of unstable particles
\begin{equation}
P_\text{decay}(\tau) = - \dv \tau \exp\left(- \Gamma \tau\right) = \Gamma \exp\left(- \Gamma \tau\right) \,.
\end{equation}
as function of the decay width $\Gamma = \Gamma(m,\theta)$.
Therefore, the total probability for an unstable and oscillating particle to decay in a proper time window is given by
\begin{equation} \label{eq:probability displacement}
P_{ll}^{\nicefrac{\LNC}{\LNV}}(\tau_{\min}, \tau_{\max}) = \int_{\tau_{\min}}^{\tau_{\max}} P^{\nicefrac{\LNC}{\LNV}}_\text{osc}(\tau) P_\text{decay}(\tau) \d \tau \,,
\end{equation}
and the number of expected events in a collider experiment is then
\begin{equation} \label{eq:event numbers}
N^{\nicefrac{\LNC}{\LNV}} = \mathscr L \sigma \BR \int D(\vartheta, \gamma) P^{\nicefrac{\LNC}{\LNV}}_{ll}(\tau_{\min}(\vartheta, \gamma), \tau_{\max}(\vartheta, \gamma)) \d\vartheta \d\gamma\,,
\end{equation}
where the factor $D(\vartheta, \gamma)$ accounts for the probability density that the heavy neutrino has Lorentz factor $\gamma$ and is produced with an angle $\vartheta$ with respect to the beam axis, $\mathscr L$ is the luminosity of the collider, and $\sigma$ and $\BR$ are the sterile neutrino production cross section and branching ratio of the process under consideration.
The parameters $\tau_{\min}$ and $\tau_{\max}$ are defined by the detector geometry when transitioning from proper time coordinates to the lab frame via $\tau(\vartheta,\gamma) = (\gamma^2-1)^{-\nicefrac12} L(\vartheta)$.

\subsection{Integrated effect} \label{sec:integrated effect}

In an experimental situation in which it is not possible to resolve the oscillation patterns, an integrated effect could still be measurable by comparing the number of events with opposite-, and same-sign leptons originating from processes such as the one presented in \cref{fig:oscillation feynman}.
Evaluating the integral \eqref{eq:probability displacement} in order to obtain the probability of \LNC and \LNV decays between the minimal proper time $\tau_{\min}$ and the maximal proper time $\tau_{\max}$ results in the difference
\begin{equation} \label{eq:finite detector probability}
P_{ll}^{\nicefrac{\LNC}{\LNV}}(\tau_{\min}, \tau_{\max})
= \Gamma \frac{P^{\nicefrac{\LNC}{\LNV}}(\tau_{\max}) - P^{\nicefrac{\LNC}{\LNV}}(\tau_{\min})}2 \,,
\end{equation}
where the indefinite integral is given by
\begin{equation}
P^{\nicefrac{\LNC}{\LNV}}(\tau) = P(\tau, \Gamma, 0) \pm \frac{P(\tau, \Gamma_-, \lambda) + P(\tau, \Gamma_+, \lambda)}2 \,,
\end{equation}
with
\begin{align}
P(\tau, \Gamma, \lambda) &= \int e^{- \lambda - \Gamma \tau } \d \tau = - \frac{e^{- \lambda - \Gamma \tau}}{\Gamma} \,, &
\Gamma_\pm &= \Gamma \pm i \Delta m \,.
\end{align}
\begin{figure}
\begin{panels}{2}
\includetikz{Rllmin}
\caption{$R_{ll}^\text{obs}(R_{ll},\tau_{\min}/\tau_\text{osc})$ while $\tau_{\max}\to\infty$.} \label{fig:Rll tau min}
\panel
\includetikz{Rllmax}
\caption{$R_{ll}^\text{obs}(R_{ll},\tau_{\max}/\tau_\text{osc})$ while $\tau_{\min}\to0$.} \label{fig:Rll tau max}
\end{panels}
\caption[$R_{ll}$ in finite detectors]{
Impact of the finite detector size on the measurement of $R_{ll}$.
The plots show the observable $R_{ll}^\text{obs}$ as a function of $\tau/\tau_\text{osc}$ and the theoretical $R_{ll}$, calculated by taking the limits \eqref{eq:Rll limits} simultaneously.
In panel \subref{fig:Rll tau min} $\tau_{\min}$ is variable and $\tau_{\max}\to\infty$ while in panel \subref{fig:Rll tau max} $\tau_{\max}$ is variable and $\tau_{\min} \to 0$.
Note that only in the region where the contours are vertical $R_{ll}$ coincides with $R_{ll}^\text{obs}$ and that $R_{ll}^\text{obs}$ is not bound to be smaller or equal to one and even has poles in panel \subref{fig:Rll tau min}.
} \label{fig:finite R_ll}
\end{figure}
\unskip In the limit that the experiment can observe all decays from the origin to infinity and under the assumption that the parameter point under consideration does allow to neglect decoherence effects, which is equivalent to
\begin{align} \label{eq:Rll limits}
\tau_{\min} &\to 0 \,, &
\tau_{\max} &\to \infty \,, &
\lambda &\to 0 \,,
\end{align}
this expression simplifies to
\begin{equation}
P_{ll}^{\nicefrac{\LNC}{\LNV}} = \frac12
\begin{cases}
\frac{\Gamma^2}{\Delta m^2 + \Gamma^2} + 1 & \text{for \LNC} \\
\frac{\Delta m^2}{\Delta m^2 + \Gamma^2} & \text{for \LNV}
\end{cases}
\,.
\end{equation}
Therefore, the ratio between the two decay modes is, in this case, given by
\begin{equation} \label{eq:ratio}
R_{ll} =
\frac{P_{ll}^{\LNV}}{P_{ll}^{\LNC}} =
\frac{\Delta m^2}{\Delta m^2 + 2 \Gamma^2} \,,
\end{equation}
which matches the result of \ccite{Anamiati:2016uxp,Das:2017hmg}.
While a Dirac heavy neutrino would have $R_{ll} = 0$ and a Majorana heavy neutrino would have $R_{ll} = 1$, a realistic pseudo-Dirac heavy neutrino can have any value in between.
However, even in the simplified case of a single Majorana heavy neutrino, the measurement of $R_{ll}=1$ would be challenging since asymmetries in the number of measured \LNC and \LNV events caused by the detector geometry in combination with the angular dependence discussed in \cref{sec:spin correlation} as well as the detector effects discussed in the following will have an impact on the measured $R_{ll}$.
For pseudo-Dirac heavy neutrinos, \cref{fig:Rll} compares the dependence of $R_{ll}$ on $\flatfrac{\Delta m}{\Gamma}$ derived in \eqref{eq:ratio} with a \MC simulation.
Additionally, the bands of $R_{ll} \in [0.1,0.9]$ for five pseudo-Dirac \BMs introduced in \cref{tab:benchmark models} are given in \cref{fig:Rll bands}.

However, taking the finite size of the experimental setup into account by integrating only over the fiducial detector size in \eqref{eq:finite detector probability} changes the picture drastically.
In this case the parameters $\tau_{\min}$ and $\tau_{\max}$ in
\begin{equation} \label{eq:observable ratio}
R_{ll}^\text{obs}(\tau_{\min}, \tau_{\max}) =
\frac{P_{ll}^{\LNV}(\tau_{\min}, \tau_{\max})}{P_{ll}^{\LNC}(\tau_{\min}, \tau_{\max})} \,,
\end{equation}
cannot be taken to be zero and infinity, respectively.
The consequences of this effect in the limit of vanishing damping and in terms of the proper time are presented in \cref{fig:finite R_ll}.
In this representation, vertical contour lines indicate negligible deviation of $R_{ll}^\text{obs}$ from $R_{ll}$.
For a finite $\tau_{\min}$, an oscillatory pattern becomes apparent that has its minima for cuts in the vicinity of multiples of the oscillation period.
The dependence on $\tau_{\max}$ is less severe and only occurs close to an $R_{ll}$ of one.
The two extreme limiting cases are
\begin{align}
R_{ll}^\text{obs} &=
\begin{dcases}
\tan^2{\frac{r^{}_{\min}}{2}} & \text{for } R_{ll} \to 0 \text{ and } \tau_{\max} \to \infty \\
\frac{r^{}_{\max} - \sin r^{}_{\max}}{r^{}_{\max} + \sin r^{}_{\max}} & \text{for } R_{ll} \to 1 \text{ and } \tau_{\min} \to 0
\end{dcases} \,, &
r^{}_m &= \Delta m \tau_m \,.
\end{align}
While the effect for $R_{ll} \to 1$ and $\tau_{\min} \to 0$ is damped for large $\tau_{\max}$ the effect for $R_{ll} \to 0$ and $\tau_{\max} \to \infty$ remains undamped when varying $\tau_{\min}$ as long as the Lorentz factor distribution, and the decoherence can be neglected \cite{Antusch:2023nqd}.

\subsection{Oscillations in the lab frame} \label{sec:lab frame}

\begin{figure}
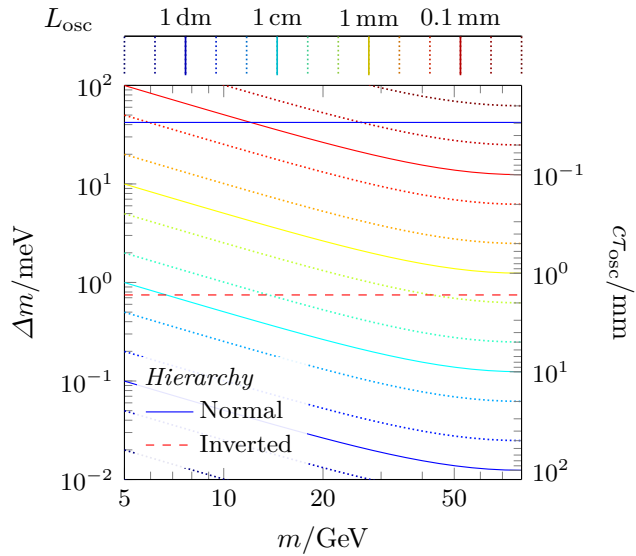

\begin{panels}{.55}
\includetikz{oscillation-length}
\end{panels}
\caption[Experimentally reachable oscillation periods]{
Oscillation period $L_\text{osc}$ as a function of the heavy neutrino mass $m$ and the mass splitting $\Delta m$.
The Lorentz factor is estimated using $W$ bosons at rest \eqref{eq:boost estimate}.
The horizontal lines correspond to the mass splitting appearing in the two minimal linear seesaw $\BMs$ given in \cref{tab:benchmark models}.
} \label{fig:oscillation length}
\end{figure}

When boosting the oscillations described in \cref{sec:lo oscillations} into the lab frame, two main effects have to be considered.
Firstly, the oscillation length in the lab frame, defined as
\begin{equation}
L_\text{osc} = \sqrt{\gamma^2 -1} \tau_\text{osc}\,,
\end{equation}
is, on an event-per-event basis, increased by the Lorentz factor.
This well-known effect
helps to compensate for the short oscillation period of some realistic \BMs.
For processes such as the one depicted in \cref{fig:oscillation feynman}, the Lorentz factor of the heavy neutrino $\gamma$ can be estimated to be
\begin{equation} \label{eq:boost estimate}
\gamma \approx \frac{m_W^2 + m^2}{2 m_W^{} m^{}}\,,
\end{equation}
which relies on the assumption that the initial $W$ boson decays at rest and neglects the mass of the prompt muon.
The resulting resolvable oscillation length as a function of the heavy neutrino mass $m$ and the mass splitting $\Delta m$ is shown in \cref{fig:oscillation length}.

Secondly, the oscillation pattern is washed out by the event-dependent Lorentz factor.
Therefore, any study hoping to resolve \NNOs needs to measure enough observables to reconstruct the Lorentz factor.
This makes studies relying on processes with final state neutrinos extremely challenging and is the reason why we focus on semileptonic final states as shown in \cref{fig:oscillation feynman}.

\section{\sentence\SPSSlong} \label{sec:SPSS}

As detectable heavy neutrinos cannot be too heavy or too weakly coupled, they must originate from a model with an approximate \LNLS and form pseudo-Dirac pairs as discussed in \cref{sec:seesaw}.
In the following, we assume that one pseudo-Dirac pair dominates the collider phenomenology.
We will first introduce the type I seesaw in the limit of exact symmetry conservation in \cref{sec:symmetric seesaw}.
This model builds on two Majorana \DOFs that form one exact Dirac particle.
Subsequently, we add small symmetry-breaking terms in \cref{sec:small breaking} recovering models such as the linear and inverse seesaw.
After comparing this model with the \LO terms governing the \NNOs summarised in \cref{sec:lo oscillations}, we introduce a phenomenological model with a minimal number of parameters that is able to describe the relevant observables in \cref{sec:pSPSS}.

\subsection{Seesaw in the symmetric limit} \label{sec:symmetric seesaw}

In the symmetric limit of the \SPSS \cite{Antusch:2015mia,Antusch:2016ejd}, the two heavy Majorana neutrinos are protected by a \LNLS.
One simple choice of charges for the \LNLS protecting the lepton number $L$ is given by
\begin{equation} \label{eq:charges}
\begin{array}{*4c} \toprule
& \ell & N_1 & N_2 \\ \midrule
L & +1 & -1 & +1 \\
\bottomrule\end{array}
\,,
\end{equation}
with all other fields having a charge of zero.
In this case, the general Lagrangian \eqref{eq:generic seesaw} takes the form
\begin{equation} \label{eq:symmetric Lagrangian}
\mathcal L_{\SPSS}^L =
\widebar{N_i^c} \ispd N_i^{}
- y_{1\alpha} \widebar{N_1^c} \widetilde H^\dagger \ell_\alpha
- \widebar{N_1^c} m_M^{} N_2^{}
+ \dots + \HC \,,
\end{equation}
where $N_1$ and $N_2$ are taken to be left-chiral sterile neutrinos.
The ellipses capture contributions from additional sterile neutrinos, which are assumed to be heavier or much weaker coupled than the explicitly denoted pair and are expected to contribute only sub-dominantly to the collider phenomenology of the model.
After \EWSB the neutrino mass Lagrangian of the interaction eigenstates $n = \row{\nu_e, \nu_\mu, \nu_\tau, N_1, N_2}^\trans$ can be written as
\begin{equation}
\mathcal L_\text{mass} = - \frac12 \widebar{n^c} M_n n + \HC \,,
\end{equation}
where the mass matrix is given by
\begin{equation}
M_n =
\begin{pmatrix}
0 & M_{3\times2} \\
M_{3\times2}^\trans & M_N
\end{pmatrix}
=
\begin{pmatrix}
0 & \vec m_D^{} & 0 \\
\vec m_D^\trans & 0 & m_M^{} \\
0 & m_M^{} & 0
\end{pmatrix}
\,,
\end{equation}
with $\vec m_D^{} = \vec y_1 v$ being the Dirac mass term.
The mass matrix can be approximately diagonalised using a Takagi decomposition
\begin{align} \label{eq:neutrino diagonlisation}
D_n &= U_n^\trans M_n U_n \,, &
U^\dagger U = \mathbb 1 \,,
\end{align}
following the steps presented in \cite{Antusch:2009gn, Antusch:2020pnn}.
Since the \LNLS is conserved, the light neutrinos are massless, and the heavy neutrinos are mass degenerate \cite{Wyler:1982dd}.
Up to the second order in the active-sterile mixing parameter
\begin{equation} \label{eq:active-sterile mixing}
\vec \theta = \frac{\vec m_D^{}}{m_M^{}},
\end{equation}
where $\vec \theta = \row{\theta_e,\theta_\mu,\theta_\tau}^\trans$,
their masses are
\begin{equation}\label{eq:mass degeneracy}
m_4 = m_5 = m_M^{} \left(1 + \frac12 \abs{\vec \theta}^2\right) + \order*{\abs{\vec \theta}^4} \,,
\end{equation}
and the mixing matrix is given by
\begin{equation}
U_n =
\begin{pmatrix}
U_{\PMNS} & U_{\CL} \\
U_{2\times 3} & U_N
\end{pmatrix}
=
\begin{pmatrix}
\mathbb 1_{3\times 3} - \frac12 \vec \theta^* \otimes \vec \theta & -\frac{i}{\sqrt2} \vec \theta^* & \frac{1}{\sqrt2} \vec \theta^* \\
0 & \frac{i}{\sqrt2} & \frac{1}{\sqrt2} \\
- \vec \theta^\trans & -\frac{i}{\sqrt2}(1 - \frac12 \abs{\vec \theta}^2) & \frac{1}{\sqrt2}(1 - \frac12 \abs{\vec \theta}^2)
\end{pmatrix}
\,,
\end{equation}
where the upper left block is the \PMNS matrix.
One can show that the upper right $3\times2$ \CL block of the mixing matrix, which relates heavy neutrinos and active \SM neutrinos, has the exact form
\begin{equation}
U_{\CL} = \frac{1}{\sqrt{1 + \abs{\vec \theta}^2}} \row*{-\frac{i}{\sqrt2} \vec \theta^*, \frac{1}{\sqrt2} \vec \theta^*} \,.
\end{equation}
Therefore it is possible to absorb higher order corrections in $\vec \theta$ into a rescaling of the Yukawa coupling.
In particular the exact form of $U_{\CL}$ can be recovered from the \LO approximation by rescaling of the mixing parameter
\begin{equation}\label{eq:rescaling}
\vec \theta^* \rightarrow \vec \theta'^* = \frac{\vec \theta^*}{\sqrt{1 + \abs{\vec \theta}^2}}\,.
\end{equation}
Consequently, it suffices to expand $U_{\CL}$ to \LO.

Due to the mass degeneracy and phase difference of the heavy neutrinos, there are no \NNOs in the symmetric limit, and this theory of two Majorana particles is equivalent to a theory with a single Dirac particle.
For any process with \SM external particles that contains a Feynman diagram with one of the heavy neutrino mass eigenstates, also the other heavy neutrino mass eigenstate contributes.
In an explicit calculation, it can be shown that the amplitudes from the different mass eigenstates cancel each other, such that \LNV processes are not present, see also \cref{sec:HNL characterisation}.
We recover the argument laid out in \cref{sec:seesaw} that a theory relying solely on one exact Dirac heavy neutrino is \LNC and incapable of generating \SM neutrino masses.

\subsection{Seesaw with small symmetry breaking} \label{sec:small breaking}

In order to introduce \SM neutrino masses, the symmetric limit of the \SPSS can be perturbed by extending the Lagrangian \eqref{eq:symmetric Lagrangian} with additional small \LNV terms
\begin{equation} \label{eq:broken Lagrangian}
\mathcal L_{\SPSS}^{\cancel L} =
- y_{2\alpha} \widebar{N_2^c} \widetilde H^\dagger \ell_\alpha
- \mu_M^\prime \widebar{N_1^c} N_1^{}
- \mu_M^{} \widebar{N_2^c} N_2^{}
+ \dots + \HC \,,
\end{equation}
where a Yukawa coupling $\vec y_2$ and two Majorana masses $\mu_M^{}$ and $\mu_M^\prime$ are introduced.
The resulting mass matrix is
\begin{equation} \label{eq:broken mass matrix}
 M_n =
\begin{pmatrix}
 0 & \vec m_D^{} & \vec \mu_D^{} \\
 \vec m_D^\trans & \mu_M^\prime & m_M^{} \\
 \vec \mu_D^\trans & m_M^{} & \mu_M^{}
\end{pmatrix}
\,,
\end{equation}
where $\vec \mu_D^{} = \vec y_2 v$ is a Dirac mass term.
These additional terms break the \LNLS symmetry characterised by the charges given in \cref{eq:charges}.
In order to ensure a small breaking of the \LNLS, we require
\begin{equation}
\order*{\frac{\vec \mu_D^{}}{m_M^{}}} = \order*{\frac{\mu_M^{}}{m_M^{}}} = \order*{\frac{\mu_M^\prime}{m_M^{}}} = \epsilon \ll 1\,,
\end{equation}
Since the parameters $\vec \theta$ can be larger than $\order \epsilon$, an expansion in small parameters might contain powers in $\vec \theta$ higher than in $\epsilon$.
The Takagi decomposition \eqref{eq:neutrino diagonlisation} of the mass matrix \eqref{eq:broken mass matrix} results now in two distinct heavy neutrino masses of
\begin{align} \label{eq:mass splitting}
m_{\nicefrac{4}{5}}^{} &= m_M^{} \left (1 + \frac12 \abs{\vec \theta}^2 \right) \mp \left(\cos(\phi) \abs*{\vec \mu_D^* \vec \theta} + \frac{\abs*{\mu_M^* + \mu_M^\prime}}2 \right) + \order*{\max(\epsilon^2, \epsilon \abs{\vec \theta}^2, \abs{\vec \theta}^4)} \,.
\end{align}
where
\begin{equation}
\phi =
\begin{cases}
0 & \text{for } \mu_M^{} = \mu_M^\prime = 0 \\
\arg(\vec \mu_D^* \vec \theta (\mu_M^* + \mu_M^\prime)) & \text{otherwise}
\end{cases}
\,.
\end{equation}
The upper right $2\times3$ part of the mixing matrix is given by
\begin{equation}
U_{\CL} =
\row*{\frac{i}{\sqrt2} \left(\vec \theta^* - \vec \theta_\epsilon^*\right),
\frac{1}{\sqrt2} \left( \vec \theta^* + \vec \theta_\epsilon^*\right)
}
+ \order*{\max(\epsilon \vec \theta, \abs{\vec\theta}^2 \vec \theta)} \,,
\end{equation}
where $\vec \theta_\epsilon = \vec \mu_D^{} / m_M^{}$ is the \LNV equivalent to the active-sterile mixing parameter.
In cases where higher order terms in $\vec\theta$ are of the same order as terms of $\order{\epsilon\vec\theta}$, the rescaling \eqref{eq:rescaling} can be used to absorb higher order terms into a redefinition of the coupling $\vec \theta$.
This shows that higher order terms in $\vec \theta$ yield no qualitative new effects, and the expansion of the mixing matrix is valid up to $\order{\epsilon\vec\theta}$.

\begin{table}
\begin{tabular}{lllr@{${}={}$}l} \toprule
\multicolumn{2}{l}{Seesaw} & Hierarchy & \multicolumn{2}{c}{\BM} \\ \midrule
\multirow{2}{*}{Linear} & \multirow{2}{*}{$\Delta m = \Delta m_\nu$} & Normal & $\Delta m_\nu$ & $\unit[(41.46\pm0.29)]{meV}$
\\
 & & Inverted & $\Delta m_\nu$ & $\unit[(749\pm21)]{\mu eV}$
\\ \cmidrule{3-5}
\multirow{3}{*}{Inverse} & \multirow{3}{*}{$\Delta m = m_\nu \abs{\vec \theta}^{-2}$} & & $m_\nu$ & $\unit[0.5]{meV}$ \\
 & & & $m_\nu$ & $\unit[5]{meV}$ \\
 & & & $m_\nu$ & $\unit[50]{meV}$ \\
\bottomrule \end{tabular}
\caption[Benchmark models]{
Five \BMs for the linear and inverse seesaw are discussed in the text.
The linear seesaw \BMs represent the only two possible models in the minimal linear seesaw of a single pseudo-Dirac heavy neutrino.
With more pseudo-Dirac heavy neutrinos in the spectrum, other models become feasible.
Since the inverse seesaw requires at least two pairs of pseudo-Dirac neutrinos, it is not possible to uniquely connect it to the measured neutrino mass differences, and we use a wide spectrum of neutrino masses as \BMs.
} \label{tab:benchmark models}
\end{table}

\begin{figure}
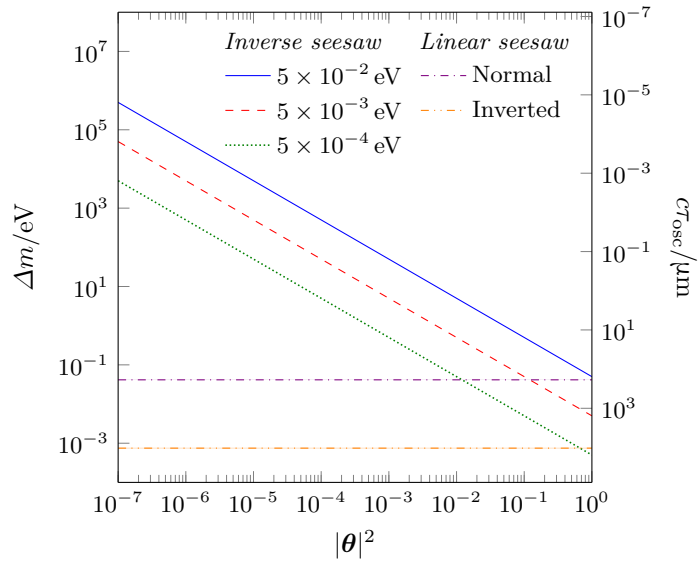

\begin{panels}{.6}
\includetikz{mass-splitting}
\end{panels}
\caption[Model dependent sterile neutrino mass splitting]{
Heavy neutrino mass splitting as a function of the active-sterile mixing parameter for the five \BMs of the minimal linear and inverse seesaw given in \cref{tab:benchmark models}.
} \label{fig:mass-splitting}
\end{figure}

The mass matrix \eqref{eq:broken mass matrix} is the most generic tree-level seesaw matrix one can generate from two sterile neutrinos.
It incorporates as limiting cases the linear and the inverse seesaw.
In particular, the individual matrix elements have the following phenomenology:
\begin{itemize}

\item
The Dirac contribution $\vec \mu_D^{}$ yields the linear seesaw \cite{Akhmedov:1995ip, Akhmedov:1995vm} for which the neutrino mass matrix is given by $M_\nu = \vec \mu_D^{} \otimes \vec \theta + \vec \theta \otimes \vec \mu_D^{}$ and has the two nonzero eigenvalues $m_\nu = \abs{\vec \mu_D^{}} \abs{\vec \theta} \mp \abs{\vec \mu_D^* \vec \theta}$.
The mass splitting of the heavy neutrinos is, in this case, $\Delta m = 2 \abs{\vec \mu_D^* \vec \theta}$.
When reproducing the measured light neutrino data in the minimal linear seesaw with a single pseudo-Dirac pair, the mass splitting of the heavy neutrinos is therefore equal to the mass splitting of the light neutrinos $\Delta m = \Delta m_\nu$, \cf \cite{Antusch:2017ebe}.

\item The Majorana contribution $\mu_M^{}$ in the third entry on the main diagonal yields the inverse seesaw \cite{Nandi:1985uh, Mohapatra:1986bd, Mohapatra:1986aw} for which the neutrino mass matrix is $M_\nu = \mu_M^{} \vec \theta \otimes \vec \theta$.
Since this expression results in a single finite eigenvalue of $m_\nu = \mu_M^{} \abs{\vec \theta}^2$ at least two pseudo-Dirac pairs are necessary to describe the observed neutrino oscillation data.
The mass splitting for the pseudo-Dirac pair considered here is $\Delta m = \abs{\mu_M^{}}$ and can therefore be expressed in terms of the \SM neutrino mass eigenvalue $\Delta m = m_\nu \abs{\vec \theta}^{-2}$, \cf \cite{Antusch:2017ebe}.

\item The Majorana contribution $\mu_M^\prime$ in the second entry on the main diagonal introduces a mass splitting between the heavy neutrinos of $\Delta m = \abs{\mu_M^\prime}$ without a \LO tree-level impact on the light neutrino masses.
It does, however, generate a contribution to $M_\nu$ at one loop \cite{Pilaftsis:1991ug,Pilaftsis:2005rv}.
Furthermore, it can be exploited to reduce the mass splitting necessary to generate realistic neutrino masses when added to the linear or inverse seesaw.
\footnote{
If additionally, the linear seesaw term is present, the subdominant contribution $\Delta M_\nu = \mu_M^\prime \vec \theta_\epsilon \otimes \vec \theta_\epsilon$ to the neutrino mass matrix is generated \cite{Lopez-Pavon:2015cga}.
}

\item The $3\times 3$ block in the upper left, which contains the light neutrino masses, is zero at tree-level.

\end{itemize}
From the observed neutrino oscillation data, it is possible to extract two mass differences which can be combined in two different mass hierarchies, the normal and inverted light neutrino mass ordering.
Whether the third neutrino is also massive cannot be determined with current data; hence we use as minimal \BMs for the linear seesaw the two possible mass differences between the two massive \SM neutrinos $\Delta m_\nu^{\text{normal}} = \unit[(41.46\pm0.29)]{meV}$ and $\Delta m_\nu^{\text{inverted}} = \unit[(749\pm21)]{\mu eV}$ \cite{Esteban:2020cvm,NuFIT:2022}.
At the same time, an upper bound on the sum of the neutrino masses $\sum m_\nu < \unit[120]{meV}$ has been established using cosmological observations \cite{Planck:2018vyg}.
Since models relying on the inverse seesaw require at least two pseudo-Dirac pairs, the mass splitting of the heavy neutrinos is not uniquely determined by the observed light neutrino parameters.
However, when defining masses for the light neutrinos as \BMs, the heavy neutrino masses can be deduced.
We use as corresponding \BMs light neutrino masses of $m_\nu = \unit[0.5,\,5,\,50]{meV}$.
These \BMs are summarised in \cref{tab:benchmark models}.

The relation between the mass splitting and the mixing parameter for the five \BMs of the linear and inverse seesaw is visualised in \cref{fig:mass-splitting}.
For \BMs with a single \LNV entry in the mass matrix \eqref{eq:broken mass matrix}, the minimal value for the mass splitting of $\Delta m \approx \unit[750]{\mu eV}$ corresponding to a maximal oscillation period of $c\tau \approx \unit[1]{cm}$ is reached in the linear seesaw with inverted ordering.
In order to generate pseudo-Dirac pairs with a smaller mass splitting and, therefore, larger oscillation period, one can, on the one hand, rely on cancellations between the linear and inverse seesaw terms or cancellations involving $\mu_M^\prime$ or, on the other hand, go beyond the minimal model consisting of a single pseudo-Dirac pair.

\subsection{\sentence\pSPSSlong} \label{sec:pSPSS}

Given the small amount of symmetry breaking permissible for consistent low-scale seesaw models, most of their effects can be neglected in phenomenological studies \cite{Kersten:2007vk}.
One important exception are \NNOs, as they are an interference phenomenon and, as such, strongly enhanced beyond the typical expectation for an $\order \epsilon$ effect.
As shown in \cref{sec:lo oscillations} the \NNOs at \LO can be described with just two parameters in addition to the symmetric limit of the \SPSS, which are the mass splitting of the heavy neutrinos $\Delta m$ governing oscillation period in the proper time frame and the damping parameter $\lambda$ governing the effects of decoherence.

Therefore, we introduce the \emph{\pSPSS} optimised for phenomenological studies.
Instead of adding the symmetry-breaking terms from Lagrangian \eqref{eq:broken Lagrangian} to the \SPSS Lagrangian \eqref{eq:symmetric Lagrangian}, one can directly add the mass splitting without specifying if it originates from a linear seesaw, an inverse seesaw, or a more complicated model.
The masses of the two heavy neutrinos in the \pSPSS are derived from the mass  in the symmetric limit and the single additional parameter $\Delta m$
\begin{equation} \label{eq:Delta m mass splitting}
m_{\nicefrac45}^{} = m_M^{} \left(1 + \frac12 \abs{\vec \theta}^2\right) \mp \frac12 \Delta m \,,
\end{equation}
capturing the entirety of sources contributing to the mass splitting in \eqref{eq:mass splitting}.
Since oscillations are an interference phenomenon, they are observable even though the mass splitting is orders of magnitude below the energies that can be hoped to be resolved in experimental setups.

Additionally, the value of the damping parameter $\lambda$ has to be determined for each parameter point.
As shown in \cite{Beuthe:2001rc,Antusch:2020pnn,Antusch:2023nqd}, the effects of decoherence described as observability conditions, and therefore the damping parameter $\lambda$, can be neglected for the part of the parameter space where it is likely that \NNOs can be reconstructed.
In order to allow the simulation of parameter points for which decoherence effects are important, we include the damping parameter in the \pSPSS.
The relevant parameters of the \pSPSS are thus the three active-sterile mixing parameters $\vec \theta$, the heavy neutrino Majorana mass $m_M^{}$, its mass splitting parameters $\Delta m$, and the damping parameter $\lambda$.

\section{Software implementation} \label{sec:implementation}

In order to simulate \NNO at colliders, we have implemented the \pSPSS in \software{FeynRules}.
We provide this implementation online \cite{FR:pSPSS} and explain its details in \cref{sec:FeynRules}.
Furthermore, we have patched \software{MadGraph} as described in \cref{sec:MadGraph} in order to be able to simulate \NNOs.

\subsection{\software{FeynRules} model file} \label{sec:FeynRules}

\begin{table}
\begin{tabular}{*3l} \toprule
\multicolumn{3}{l}{\texttm{BLOCK PSPSS \#}} \\\midrule
\code{1} &\code{1.000000e+02} &\code{# mmaj}\\
\code{2} &\code{1.000000e-12} &\code{# deltam}\\
\code{3} &\code{0.000000e+00} &\code{# theta1}\\
\code{4} &\code{1.000000e-03} &\code{# theta2}\\
\code{5} &\code{0.000000e+00} &\code{# theta3}\\
\code{6} &\code{0.000000e+00} &\code{# damping}\\
\bottomrule\end{tabular}
\caption[\software{FeynRules} model file parameter]{
The parameters of the \pSPSS implemented in the \software{FeynRules} model file as they appear in the \software{MadGraph} \nolinkurl{param.card}.
} \label{tab:parameter}
\end{table}

In the following, we introduce the quantities used in the \software{FeynRules} \cite{Alloul:2013bka} model file \cite{FR:pSPSS}.
For each quantity we give, besides its mathematical symbol, also its variable name used in the model file, denoted in brackets.

According to the discussion in the previous section, the \SM Lagrangian has to be extended by two sterile Majorana neutrinos denoted by $N_1$ (\code{N1L}) and $N_2$ (\code{N2L}).
In the model file, an additional \NP parameter for the Higgs \VEV denoted by \code{vevNP} has been introduced.
The only difference to the usual Higgs \VEV (\code{vev}) is the interaction order, which has been set to \code{{NP, -1}} instead of \code{{QED, -1}} in order to provide \software{MadGraph} with the correct power counting rules, but for all physical considerations they can be thought of as equal.
Along with \code{vevNP}, the field \code{PhiNP} is introduced, which corresponds to the usual Higgs doublet \code{Phi} where the \code{vev} is replaced by \code{vevNP}.
The masses of the heavy neutrinos are given by \eqref{eq:Delta m mass splitting} where the Majorana mass parameter is denoted by $m_M^{}$ (\code{Mmaj}), the active-sterile mixing parameter given in \eqref{eq:active-sterile mixing} by $\vec \theta$ (\code{theta1}, \code{theta2}, \code{theta3}), and the mass splitting of the heavy neutrinos is implemented via the parameter $\Delta m$ (\code{deltaM}).

The physical fields, \ie the mass eigenstates, are extended by two self-conjugate neutrinos labelled $n_4$ (\code{n4}) and $n_5$ (\code{n5}).
The mixing matrix relating the mass eigenstate neutrinos to the interaction eigenstates is denoted by $U_n$ (\code{Un}) and satisfies
\begin{equation}
U_n^\trans M_n U_n = \diag(0, 0, 0, m_4^{}, m_5^{}) \,,
\end{equation}
up to second order in $\vec \theta$.
The matrix $U_{\CL}^\prime$ (\code{UnCL}) is the upper right $3\times5$ part of $U_n$ and is necessary for the automatic index contraction in \software{FeynRules}.
A field that contains all mass eigenstate neutrinos is introduced as $n_i$ (\code{nL}).
With this, the active neutrino interaction eigenstates of the \SM can be rotated to the mass eigenstate neutrinos using the relation
\begin{equation} \label{eq:neutrino rotation}
n_\alpha = (U_{\CL}^\prime)_{\alpha i} n_i \,.
\end{equation}
The neutrinos interact with the \SM only via the lepton doublet $\ell$ (\code{LL}).
The mixing is implemented using \cref{eq:neutrino rotation} by replacing the neutrino $\nu$ (\code{vl}) with the transformed neutrino (\code{UnCL nL}).

The kinetic terms of the sterile neutrinos
\begin{verbatim}
I (N1Lbar.Ga[mu].del[N1L, mu] + N2Lbar.Ga[mu].del[N2L, mu])
\end{verbatim}
have been added to the \code{LFermions} Lagrangian of the \SM.
The mass term
\begin{verbatim}
- Mmaj CC[N1Lbar[sp1]].N2L[sp1]
\end{verbatim}
where \code{sp1} is an internal spin index and the Yukawa term
\begin{verbatim}
yvn[ff1] (CC[N1Lbar[sp1]].LL[sp1, ii, ff1] PhiNPbar[jj] Eps[ii, jj])
\end{verbatim}
where \code{yvn} is the Yukawa vector $\vec y_{\alpha1}$, \code{ff1} is an internal family index and \code{ii} and \code{jj} are internal $\SU(2)$ doublet indices, have been introduced in the \NP Lagrangian (\code{LNP}).
Finally, the complete Lagrangian describing \SM interactions of the \pSPSS is denoted by \code{LpSPSS}.
Since the implemented mixing matrix $U_n$ is only valid up to the second order in $\vec \theta$, the full Lagrangian must be expanded in $\vec \theta$ and terms smaller than $\order{\abs{\vec \theta}^2}$ have to be neglected.
This is achieved using the function \code{RemoveHigherOrder} defined in the model file that expands its arguments up to second order in $\vec \theta$ and is automatically applied to the relevant Lagrangian parts.

The only additional free parameters in this model collected in the \code{PSPSS} block are the Majorana mass $m_M^{}$ (\code{Mmaj}), the mass splitting $\Delta m$ (\code{deltaM}), and the three active-sterile mixing parameters $\vec \theta$ (\code{theta1}, \code{theta2}, \code{theta3}).
The damping parameter $\lambda$ (\code{damping}) has also been implemented in the \software{FeynRules} model file and can be adjusted in the \software{MadGraph} \nolinkurl{param.card}.
It is set to zero by default.
When it is nonzero, it has the effect of damping the oscillations as discussed in \cref{sec:lo oscillations}.
All parameters as they appear in the \nolinkurl{param.card} are collected in \cref{tab:parameter}.

\subsection{Oscillations in \software{MadGraph}} \label{sec:MadGraph}

\begin{figure}
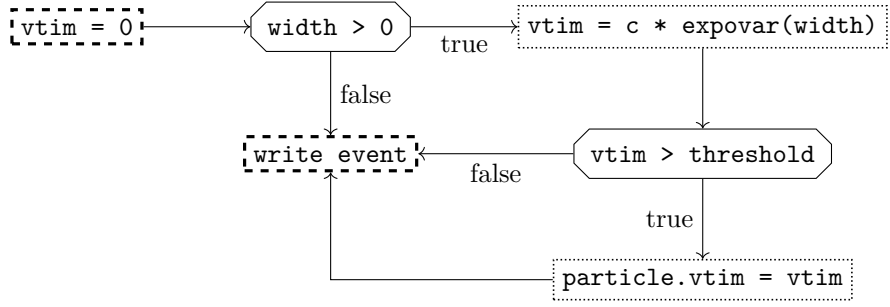

\includetikz*{flowchart_base}
\caption[Flowchart of the \software{MadGraph} \TOFlong calculation]{
The usual routine of \software{MadGraph} assigning the proper $\TOF$ (\texttm{vtim}) to a particle (\texttm{particle}) as long as it is larger than the threshold value (\texttm{threshold}) given in the \texttv{run.card}.
The process flows from the node \texttm{\lstyle vtim = 0} to the node \texttm{write event}.
} \label{fig:flowchart original}
\end{figure}

\begin{figure}
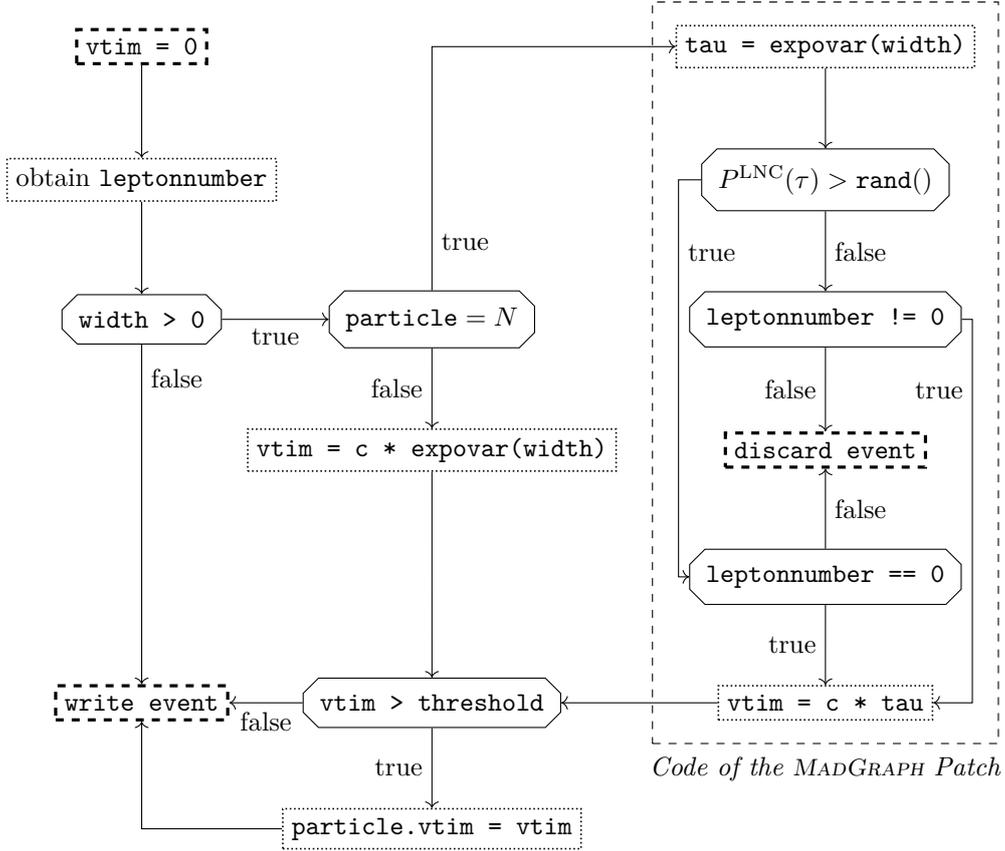

\includetikz*{flowchart_osci}

\caption[Flowchart of the patched \software{MadGraph} \TOFlong and oscillation calculation]{
Patched \software{MadGraph} routine assigning a proper $\TOF$ (\texttm{vtim}) in the presence of oscillations to a particle (\texttm{particle}).
The process flows from the node \texttm{\lstyle vtim = 0} to the nodes \texttm{write event} and \texttm{discard event}.
} \label{fig:flowchart modified}
\end{figure}

The \UFO folder exported from \software{FeynRules} can be imported into \software{MadGraph} \cite{Alwall:2014hca} using the \code{import model} command.
In order to generate the process depicted in \cref{fig:oscillation feynman}, one defines the new multi-particles
\begin{verbatim}
define mu = mu+ mu-
define ww = w+ w-
define nn = n4 n5
\end{verbatim}
As explained below, it is necessary to force the heavy neutrino to be on-shell.
This can be achieved by generating a process that contains \LNC as well as \LNV events via
\begin{verbatim}
generate p p > mu nn, (nn > mu j j)
\end{verbatim}
If the heavy neutrino mass is significantly below the $W$ mass, it is also possible to additionally ignore off-shell effects from the $W$ boson by using
\begin{verbatim}
generate p p > ww, (ww > mu nn, (nn > mu j j))
\end{verbatim}
A directory with the code for this process can then be obtained using the \code{output} command.
In the following, the path of this directory is denoted by \nolinkurl{[pSPSS]}.

There are two reasons why it is desirable to ensure that the heavy neutrino is produced on-shell.
First, the derivation of the oscillation formulae is based on the Jacob-Sachs theorem, see \cite{Beuthe:2001rc,Jacob:1961zz}, which takes the heavy neutrinos on-shell for large enough times.
In practice, the constraint to be \emph{large enough} does not give any restriction stronger than the ones already obtained in the observability conditions \cite{Antusch:2023nqd}, such that in cases where there are oscillations, the heavy neutrino can be taken to be on-shell.
For this paper, it has been assumed that decoherence can be neglected, and therefore effects from times smaller than the threshold are also neglected.
The second reason to force the heavy neutrino to be on-shell is due to the implementation details of the \software{MadGraph} patch given here.
Since parameter points that feature oscillations are close to the symmetry limit, one expects almost no \LNV events for a prompt decay.
Since \software{MadGraph} simulates the process as if it were prompt and adds displacement afterwards, one would not get \LNV events if interference between different mass eigenstates is taken into account; see also the diagrammatic explanation in \cref{sec:HNL characterisation}.
Taking the heavy neutrinos on-shell destroys the interference between different mass eigenstates such that \LNC and \LNV events are created with equal probability.
The correct interference patterns, featuring oscillations and the correct ratio of \LNC and \LNV events, are then added with the implementation of the patch given here.

The patch has been developed and tested with \software[2.9.10 (LTS)]{MadGraph5\_aMC@NLO} and is given in \cref{sec:patch}.
The idea behind it necessitates changing the behaviour of the function \code{do_add_time_of_flight} located in the file \nolinkurl{[pSPSS]/bin/internal/madevent_interface.py},
\footnote{
If one wants to patch \software{MadGraph} globally, instead of just patching the individual process, one can apply the patch to the file located in \nolinkurl{[MadGraph]/madgraph/interface/madevent_interface.py} instead.
}
which derives the decay vertex of a long-lived particle.
The main routine that is executed for each particle in each event and which assigns the \TOF to the particles is depicted in \cref{fig:flowchart original}.
This routine is changed as described in \cref{fig:flowchart modified}.
The main idea is to compare the oscillation probability at a given proper time $\tau$, to a pseudo-random variable in the range $[0, 1)$.
Based on this, it is decided if the heavy neutrino should decay in a \LNC or \LNV process.
Comparing this with the actual lepton number of the process (\code{leptonnumber}), which is zero in the \LNC case and different from zero in the \LNV case, it is then decided if the event should be kept or discarded.
This algorithm ensures that any spin correlations that are present in the generated events are kept, but the drawback is that half of all events containing heavy neutrinos are discarded.
\footnote{
A possible alternative solution that does not keep the spin correlation but does keep all events could be to change the charge of the second lepton depending on whether the event should be \LNC or \LNV.
}

In order to activate the computation of \TOF, one also has to set the parameter \code{time_of_flight} in the \code{run.card} to an appropriate threshold value (\code{threshold}) value, \eg to zero.
Events can then be generated in the usual way using \software{MadEvent}.
As described in \cref{sec:results,sec:patch} we have carefully checked that the generated data faithfully represents the physical processes.

\section{Example results} \label{sec:results}

We have calculated various observables within the \pSPSS using the model file and code modification described in \cref{sec:implementation}.
For all of the results gathered here, we have set $\theta_e = \theta_\tau = 0$ and assumed that the effects of decoherence on the \NNOs can be neglected, \ie $\lambda = 0$.
In order to demonstrate that we reproduce prior results, we show as an example the heavy neutrino decay width and the expected number of events in a \CMS-like detector as a function of the heavy neutrino mass and its coupling strength in \cref{sec:expected events}.
Furthermore, we derive the most likely Lorentz factors as well as the fraction of events with a larger Lorentz factor in \cref{sec:Lorentz factors}.
The main goal of this work is presented in \cref{sec:result oscillations} where we show an example of \NNOs.
Afterwards, the integrated effect is discussed in \cref{sec:result integrated effect}.
In \cref{sec:spin correlation}, the dependence of the transverse impact parameter on the angular-dependent spin correlation is presented.
We conclude the results in \cref{sec:bounds} by commenting on how displaced and prompt \HNLs searches can be interpreted as bounds on the pseudo-Dirac heavy neutrino of low-scale seesaw models.

\subsection{Observable events at an \LHC experiment} \label{sec:expected events}

\begin{figure}
\begin{panels}[t]{.515}
\includetikz{width}
\caption{Decay width in \unit{eV}.} \label{fig:decay width}
\panel{.485}
\includetikz{expected_events}
\caption{Number of expected events $N$.
} \label{fig:event number}
\end{panels}
\caption[Heavy neutrino decay width and number of expected events at the $\HLLHC$]{
Panel \subref{fig:decay width}: Scan over the decay width of the heavy neutrino as a function of its mass and the active-sterile mixing parameter $\abs{\vec \theta}^2$.
Panel \subref{fig:event number}: Scan over the number of expected displaced vertex events in the \CMS detector for the $\HLLHC$ with $\mathscr L = \unit[3]{\inv{ab}}$ and cuts as defined in \cref{sec:expected events}.
} \label{fig:decay width and event number}
\end{figure}

We present a scan over the parameter space resulting in the decay width and the number of expected events for a given experimental setup and luminosity.
For each parameter point, the decay width $\Gamma$ and the cross section $\sigma$ are computed, and the event obtained from \software{MadGraph} is used for a toy analysis.
We present the decay width as a function of the heavy neutrino mass $m$ and the active-sterile mixing parameter $\abs{\vec \theta}^2$ in \cref{fig:decay width}.
In order to derive the expected number of events, we employ the most important cuts used for displaced vertex analyses at the \LHC.
For this example, we use cuts inspired by the \CMS Phase II detector:
\begin{itemize}
\item Minimal transverse momentum $p_T^{\min}(\mu_\text{prompt}) = \unit[20]{GeV}$ of the prompt muon
\item Minimal transverse momentum $p_T^{\min}(f_\text{disp}) = \unit[1]{GeV}$ of the displaced muon and quarks
\item Maximal pseudorapidity $\eta^{\max} = 4$ for leptons and quarks
\item Minimal impact parameter $d_0^{\min}(\mu_\text{disp}) = \unit[2]{mm}$ of the displaced muon
\item Maximal distance of half the tracker size for displaced muon and quarks
\end{itemize}
Where the $p_T^{\min}(\mu_\text{prompt})$ cut is used to ensure that the event is triggered, the $p_T^{\min}(f_\text{disp})$ and $\eta^{\max}$ cuts ensure that the particles are captured by the detector, and the $d_0^{\min}(\mu_\text{disp})$ ensures that the secondary muon is indeed reconstructed as a displaced muon.
Finally, the cut on the maximal displaced distance is used to ensure that the tracks of the displaced muons and of the daughters of the quarks can be measured such that the displaced vertex can be reconstructed.
With the simulated events and the above-mentioned cuts, an efficiency factor and the number of expected events are computed via
\begin{align}
N_\text{exp} &= \sigma \mathscr L f_\text{eff} \,, &
f_\text{eff} &= \frac{N_\text{after cuts}}{N_\text{all events}}\,,
\end{align}
The resulting expected number of events as a function of the heavy neutrino mass and active-sterile mixing parameter are shown in \cref{fig:event number}.

\subsection{Maximal Lorentz factors} \label{sec:Lorentz factors}

\begin{figure}
\begin{panels}{.45}
\includetikz{gamma}
\caption{Example distribution of Lorentz factors.} \label{fig:Lorentz factor distribution}
\panel{.55}
\includetikz{lorentz-factors}
\caption{Resolvable mass splittings for $L_\text{osc} = \unit[2]{mm}$.} \label{fig:resolvable mass splittings}
\end{panels}
\caption[Lorentz factor dependence of the resolvable mass splitting]{
Panel \subref{fig:Lorentz factor distribution}: Histogram of the simulated Lorentz factor \PDF for heavy neutrinos with a mass of \unit[7]{GeV}.
Panel \subref{fig:resolvable mass splittings}: Resolvable mass splittings for a fixed oscillation length of $L_\text{osc} = \unit[2]{mm}$ for event samples with different Lorentz factor thresholds $\gamma^f$.
The horizontal lines correspond to the mass splitting appearing in the two minimal linear seesaw $\BMs$ given in \cref{tab:benchmark models}.
} \label{fig:Lorentz factors}
\end{figure}

\begin{figure}
\begin{panels}{2}
\includetikz{gammas}
\caption{Most likely Lorentz factors as function of $m$.} \label{fig:most likely Lorentz factors plot}
\panel
\begin{tabular}{lll} \toprule
$f$ & $\gamma_W^{}$ & $\alpha$ \\ \midrule
$1$ & $1$ & $0$ \\
$\nicefrac12$ & $3.66$ & $\expnumber{3.839}{-7}$ \\
$\nicefrac13$ & $7.92$ & $\expnumber{1.545}{-5}$ \\
$\nicefrac1{10}$ & $17.36$ & $\expnumber{6.525}{-1}$ \\
\bottomrule \end{tabular}
\vspace{7ex}
\caption{Fit values} \label{fig:tab}
\end{panels}
\caption[Fit of the most likely Lorentz factor thresholds]{
Panel \subref{fig:most likely Lorentz factors plot}: Most likely Lorentz factors $\widehat \gamma$ of heavy neutrinos as a function of their masses.
The points correspond to $\MC$ data and the lines to a fit.
Panel \subref{fig:tab}: fit parameter of the function \eqref{eq:boost estimate 2} for the most likely Lorentz factors $\widehat \gamma$ after applying a threshold $f$.
} \label{fig:most likely Lorentz factors}
\end{figure}

Based on the simulated events, the estimation of the Lorentz factor of the heavy neutrino in \cref{eq:boost estimate} can be tested.
Since this relies on the probability distributions for the Lorentz factor of the heavy neutrinos, one example of such a Lorentz factor distribution is shown in \cref{fig:Lorentz factor distribution}.
In the following, we define the most likely Lorentz factor $\widehat \gamma$ as the one where the distribution has its maximum.

It is also possible to consider only a fraction of events with the largest boost in order to probe the smaller mass splittings that become accessible since the most likely Lorentz factor will be increased in such a data sample.
We define the Lorentz factor threshold $\gamma^f$ such that only the fraction $f$ of events with the highest Lorentz factors is kept.
The mass splittings, still resolvable for different thresholds and a fixed oscillation length of $L_\text{osc} = \unit[2]{mm}$, are depicted in \cref{fig:resolvable mass splittings}.

In addition to the information in \cref{eq:boost estimate}, the boost of the initial $W$ boson can also be taken into account.
To that end, the momentum of the heavy neutrino in the rest frame of the $W$ boson is multiplied by a fitting parameter $\alpha\in[0,1]$ estimating the component parallel to the $W$ boson momentum.
Finally, the energy and parallel momentum are boosted with $- \beta_W$ from the $W$ rest frame to the lab frame, yielding
\begin{equation} \label{eq:boost estimate 2}
\gamma \approx
\gamma_W^{} \left(
\frac{m_W^2 + m^2}{2 m_W^{} m}
+ \alpha \frac{m_W^2 - m^2}{2 m_W^{} m} \sqrt{1 - \frac1{\gamma_W^2}}
\right) \,.
\end{equation}
The parameters $\alpha$ and $\gamma_W^{}$ are obtained by fitting this equation to simulated data; the results are presented in \cref{fig:most likely Lorentz factors}.
Comparing this most likely Lorentz factor to the estimate of the Lorentz factor \cref{eq:boost estimate 2} shows that $\gamma_W^{} \approx 1$ as used in \cref{eq:boost estimate} is a good approximation when considering the full event sample as can be seen in \cref{fig:most likely Lorentz factors plot}.
The values for the constrained event samples have been used in the derivation of \cref{fig:resolvable mass splittings}.

\subsection{Oscillations} \label{sec:result oscillations}

\begin{figure}
\begin{panels}{2}
\includetikz{dist}
\caption{Lab frame.} \label{fig:lab frame}
\panel
\includetikz{tau}
\caption{Proper time frame.} \label{fig:proper time frame}
\end{panels}
\caption[Heavy neutrino-antineutrino oscillations]{
\sentence\PDFs of the \NNOs in the lab frame are shown in panel \subref{fig:lab frame}, and the ones in the proper time frame are shown in panel \subref{fig:proper time frame}.
The parameter point used for these plots features heavy neutrinos with a Majorana mass of $m_M = \unit[20]{GeV}$, an active-sterile mixing of $\abs{\vec \theta}^2 = 10^{-8}$, and a mass splitting of $\Delta m = \unit[748]{\mu eV}$.
The latter corresponds to the linear seesaw \BM point with inverted hierarchy.
The oscillations are simulated by \software{MadGraph} using the model file \cite{FR:pSPSS} presented in \cref{sec:FeynRules} after applying the patch explained in \cref{sec:MadGraph} and explicitly given in \cref{sec:patch}.
While the oscillations in the lab frame are washed out, they are clearly visible in the proper time frame.
} \label{fig:pattern}
\end{figure}

The main purpose of this work is to provide the necessary tool to study \NNOs at colliders.
We use the \software{FeynRules} model file of the \pSPSS together with the \software{MadGraph} patch presented here to explore if such oscillations are potentially detectable at a \CMS-like detector.
In \cref{fig:pattern} the oscillation pattern as simulated by \software{MadGraph} is depicted in the lab frame as well as in the proper time frame.
As discussed in \cref{sec:lab frame} the oscillation pattern in the lab frame is washed out since heavy neutrinos with different Lorentz factors overlap.
However, the oscillations in the proper time frame of the heavy neutrino are not affected by the washout caused by the Lorentz factor, and the oscillation pattern is visible.
The detectability of \NNOs at the \CMS detector during the \HLLHC is studied in detail in \cite{Antusch:2022hhh}.

\subsection{Integrated effect} \label{sec:result integrated effect}

\begin{figure}
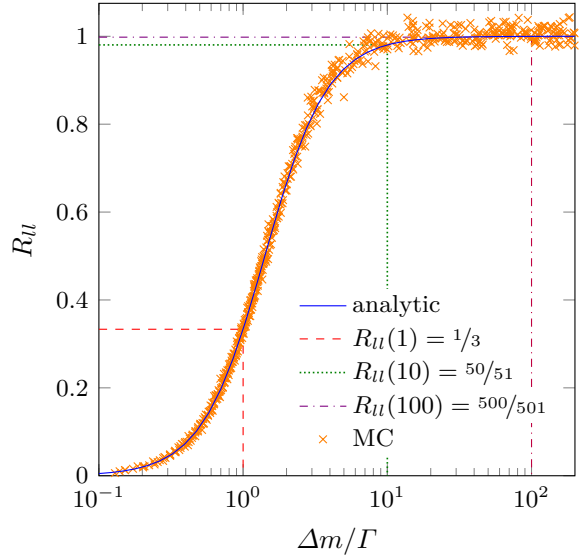

\begin{panels}{2}
\includetikz{ratio}
\end{panels}
\caption[Comparison between simulation and calculation of $R_{ll}$]{
Comparison between the analytic calculation \eqref{eq:ratio} and simulated events for the $\LNV$ over $\LNC$ ratio $R_{ll}$ as a function of the ratio $\Delta m / \Gamma$.
The three indicated special cases for $R_{ll}(\Delta m/\Gamma)$ correspond to the oscillations shown in \cref{fig:oscillations}.
} \label{fig:Rll}
\end{figure}

\begin{figure}
\begin{panels}{2}
\includetikz{Rllobsmin}
\caption{$R_{ll}^\text{obs}(R_{ll},\tau_{\min}/\tau_\text{osc})$ while $\tau_{\max}\to\infty$.} \label{Rll obs tau min}
\panel
\includetikz{Rllobsmax}
\caption{$R_{ll}^\text{obs}(R_{ll},\tau_{\max}/\tau_\text{osc})$ while $\tau_{\min}\to0$.} \label{Rll obs tau max}
\end{panels}
\caption[Simulation of $R_{ll}$ in finite detectors as function of $\tau$]{
\MC simulation of the reliability of the observation of $R_{ll}$ when considering finite detectors.
The variables are the same as in \cref{fig:finite R_ll}.
The plot in panel \subref{Rll obs tau min} agrees only in the lower right half with the theoretical results \cref{fig:Rll tau min}, since for large cut values and small $R_{ll}$ no events are generated.
The plot in panel \subref{Rll obs tau max} reproduces the analytic behaviour from \cref{fig:Rll tau max}.
} \label{fig:MC finite R_ll}
\end{figure}

\begin{figure}
\begin{panels}{3}
\includetikz{Rllobsd0min10}
\caption{$d_0$ cut with $m=\unit[10]{GeV}$.} \label{Rll obs d0 10 GeV}
\panel
\includetikz{Rllobsd0min100}
\caption{$d_0$ cut with $m=\unit[100]{GeV}$.} \label{Rll obs d0 100 GeV}
\panel
\includetikz{Rllobsd0min1000}
\caption{$d_0$ cut with $m=\unit[1]{TeV}$.} \label{Rll obs d0 1 TeV}
\panel
\includetikz{Rllobsdmin10}
\caption{$d_\text{min}$ cut with $m=\unit[10]{GeV}$.} \label{Rll obs dmin 10 GeV}
\panel
\includetikz{Rllobsdmin100}
\caption{$d_\text{min}$ cut with $m=\unit[100]{GeV}$.} \label{Rll obs dmin 100 GeV}
\panel
\includetikz{Rllobsdmin1000}
\caption{$d_\text{min}$ cut with $m=\unit[1]{TeV}$.} \label{Rll obs dmin 1 TeV}
\panel
\includetikz{Rllobsdmax10}
\caption{$d_\text{max}$ cut with $m=\unit[10]{GeV}$.} \label{Rll obs dmax 10 GeV}
\panel
\includetikz{Rllobsdmax100}
\caption{$d_\text{max}$ cut with $m=\unit[100]{GeV}$.} \label{Rll obs dmax 100 GeV}
\panel
\includetikz{Rllobsdmax1000}
\caption{$d_\text{max}$ cut with $m=\unit[1]{TeV}$.} \label{Rll obs dmax 1 TeV}
\end{panels}
\caption[Simulation of $R_{ll}$ in finite detectors as function of $d_0$, and $d$]{
Simulation of $R_{ll}$ under consideration of a finite $d_0$ in panels \subref{Rll obs d0 10 GeV}, \subref{Rll obs d0 100 GeV}, and \subref{Rll obs d0 1 TeV}, of a finite $d_{\min}$ in panels \subref{Rll obs dmin 10 GeV}, \subref{Rll obs dmin 100 GeV}, and \subref{Rll obs dmin 1 TeV} as well as of a finite $d^{\min}$ in panels \subref{Rll obs dmax 10 GeV}, \subref{Rll obs dmax 100 GeV}, and \subref{Rll obs dmax 1 TeV}.
Each cut is shown for three different masses $m= \unit[10,\,100,\,1000]{GeV}$.
} \label{fig:MC finite R_ll 2}
\end{figure}

In the cases where oscillations cannot be directly resolved, there might still be non-trivial relations between \LNC and \LNV events.
The integrated effect presented in \cref{sec:integrated effect} depends on the interplay between the decay width and the oscillation period as shown in \cref{eq:ratio}.
Counting the number of the opposite- and same-sign events while scanning the parameter space over different values for the Majorana mass $m_M^{}$ and the mass splitting $\Delta m$ reproduces the analytic dependence of \eqref{eq:ratio} as shown in \cref{fig:Rll} \cite{Anamiati:2016uxp,Drewes:2019byd,Fernandez-Martinez:2022gsu}.

Furthermore, the \MC simulation reproduces the analytic calculation leading to \cref{fig:finite R_ll} when taking the finite detector size into account as shown in \cref{fig:MC finite R_ll}.
The \MC simulation agrees with the analytic calculation for $\tau_{\min} \to 0$.
However, for $\tau_{\max} \to 0$, the upper left part of the plot is inaccessible since no events are generated for these parameter points.
In particular, this effect regularises the divergent parts in the corresponding analytic calculation.
Although the cut in $\tau$ corresponds to the cleanest theoretical description of $R_{ll}$ with finite detector effects, it does not correspond to realistic physical cuts.
Therefore, we present additional results using a $d_0$ cut, a minimal distance lab frame cut $d_{\min}$, and a maximal distance lab frame cut $d_{\max}$ in \cref{fig:MC finite R_ll 2}.
Since the boost to the lab frame introduces a dependence on the Lorentz factor, we report the results for the three mass points $m = \unit[10,\,100,\,1000]{GeV}$.

For the masses above the $W$ threshold, the $d_0$ cut reproduces approximately the first maximum of the analytic calculation.
Furthermore, the area in which no events are generated becomes larger, and for smaller masses, the first maximum is pushed to larger values of the oscillation period.
When applying a minimal distance cut in the lab frame $d_{\min}$, parts of the behaviour described for a $\tau_{\min}$ and $d_0$ cut are washed out, and the vertical lines that indicate a trivial relation between $R_{ll}$ and $R_{ll}^\text{obs}$ are recovered.
However, they are shifted and compressed towards lower values of $R_{ll}$.
Nonetheless, the maximum remains, although it is shifted to larger values of the oscillation period.
A maximal distance cut in the lab frame $d_{\max}$ does not show the oscillatory pattern and leads only to small distortions for cuts larger than the oscillation period.
However, for cuts below one to two oscillation periods, large deviations can be observed.

We conclude that although the Lorentz factor distribution smears the finite detector effect observed in \cref{sec:integrated effect}, it remains important and cannot be neglected, especially when applying a $d_0$ cut of the order of the oscillation period.

\subsection{Transverse impact parameter and spin correlation} \label{sec:spin correlation}

\begin{figure}
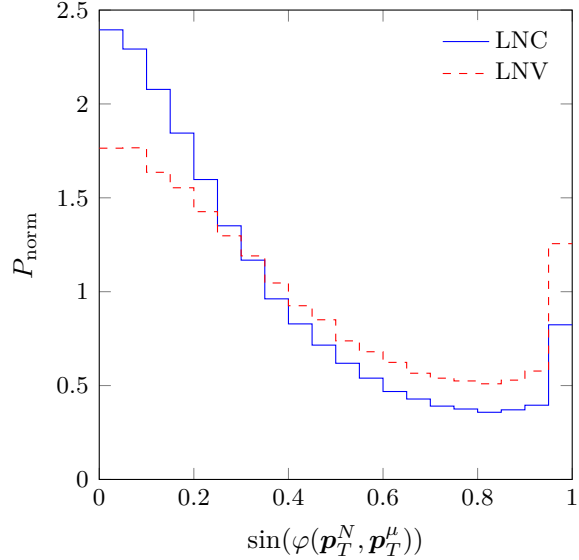

\begin{panels}{2}
\includetikz{spin-correlation}
\end{panels}
\caption[Angular dependence of heavy neutrino decays]{
Comparison between the angular dependence of the \LNC process compared to the \LNV process in the observable \eqref{eq:impact parameter simplified} using a \MC simulation.
} \label{fig:spin correlation}
\end{figure}

The transverse impact parameter $d_0$ of particles emerging in a secondary vertex is defined as
\begin{equation} \label{eq:impact parameter}
d_0 =
\frac{\vec d_T^\prime \wedge \vec p_T^\prime}{p_T^\prime} =
\frac{\epsilon_{ij} x_i^\prime p_j^\prime}{p_T^\prime} =
\frac{x_{}^\prime p_y^\prime - y_{}^\prime p_x^\prime}{p_T^\prime}
\,,
\end{equation}
where $\vec d_T^\prime = \row{x^\prime, y^\prime}$ is the position of the point with the smallest distance to the $z$-axis and $\vec p_T^\prime = \row{p_x^\prime, p_y^\prime}$ is the transverse momentum at this point.
\footnote{The two-dimensional wedge product results in a scalar quantity.}
In cases where the transverse momentum of the particle is large or the magnetic field is small, such that the radius of the trajectory of the particle is much larger than the relevant length scales of the detector, the trajectory is well approximated by a straight line.
In this case, the value of $d_0$ is constant for every point on the trajectory.
Therefore, it is possible to replace $\vec d_T^\prime$ with the coordinates of the production vertex $\vec d_T = (x,y)$, which coincides with the transverse distance of the production vertex.
In the case of the displaced muon, whose production vertex is the decay vertex of the heavy neutrino, one can furthermore use the relation
\begin{equation}
\vec d_T^\prime \rightarrow \vec d_T = d_T \frac{\vec p_T^N}{p_T^N}\,,
\end{equation}
where $\vec p_T^N$ is the transverse momentum of the heavy neutrino.
This yields a simplified definition of the impact parameter for displaced muons, under the assumption that their trajectories can be approximated as straight lines
\begin{align} \label{eq:impact parameter simplified}
d_0 &\simeq d_T \frac{\vec p_T^N\! \wedge \vec p_T^\mu}{p_T^N p_T^\mu} = d_T \sin(\varphi(\vec p_T^N, \vec p_T^\mu)) \,, &
\sin(\varphi(\vec p_T^N, \vec p_T^\mu)) &= \frac{\vec p_T^N\! \wedge \vec p_T^\mu}{p_T^N p_T^\mu} \,.
\end{align}
Therefore, the impact parameter consists, besides the transverse decay length of the invisible particle, of an angular-dependent part.
The angular-dependent part encodes a spin dependence which can be exploited to distinguish the \LNC and \LNV processes as shown in \cref{fig:spin correlation}.
One can see that just by measuring this or a related quantity such as the angle between the two charged leptons in \cref{fig:oscillation feynman}, it is possible to compute probabilities for the event being \LNC or \LNV \cite{Tastet:2019nqj}.
On the other hand, imposing a minimal $d_0$ cut is the standard strategy to reduce the background of displaced \HNL searches as seen in \cref{sec:expected events}.
The angular dependence of $d_0$ leads after cuts to a residual oscillation in the observable $N_{\LNC}+N_{\LNV}$ that naively would be expected to be free of oscillations.

\subsection{Experimental bounds on low-scale seesaw models} \label{sec:bounds}

\begin{figure}
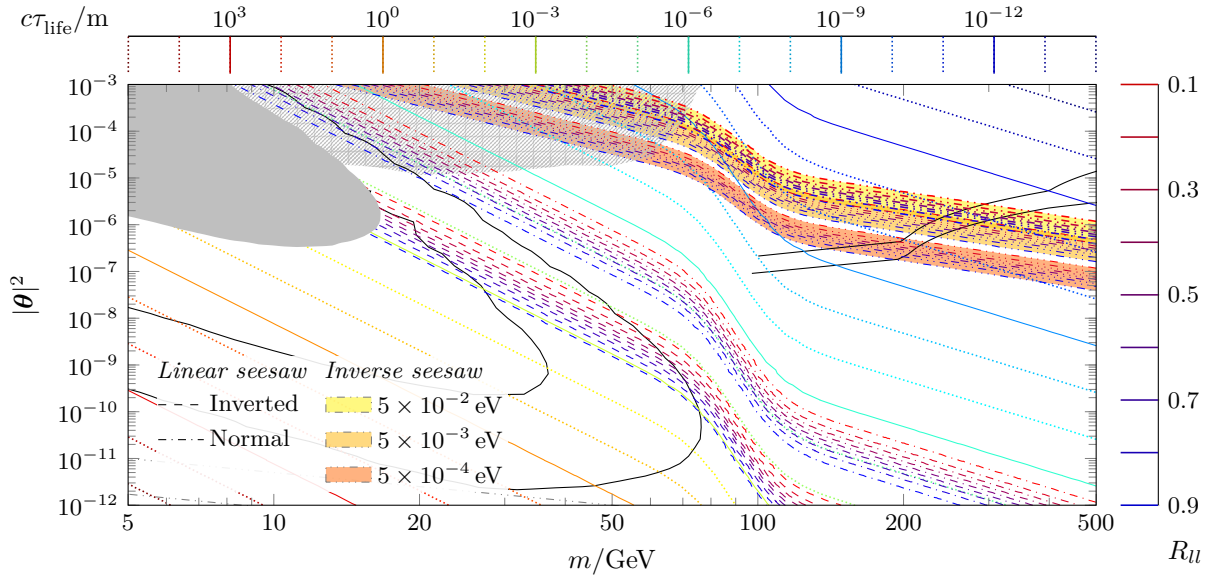

\includetikz{decaylength}
\caption[Model dependent comparison between mean lifetime and $R_{ll}$]{
Contour lines for the heavy neutrino mean lifetime $c\tau_\text{life}^{}$ in \unit{m} overlayed by contour bands of $R_{ll} \in [0.1,0.9]$ for the \BMs summarised in \cref{tab:benchmark models}.
While the two bands generated by the linear seesaw are well separated and reach far below the $W$ mass threshold, the bands generated by the inverse seesaw are very close to each other and only reach below the $W$ threshold for comparably small values of the active-sterile mixing parameter.
For comparison, the regions excluded by the \ATLAS and \CMS experiments via displaced searches are shown as grey areas \cite{ATLAS:2020xyo, CMS:2022fut}.
The areas shaded in grey at the centre top are excluded by prompt searches \cite{ATLAS:2015gtp, CMS:2018iaf}, as long as the prompt same-sign dilepton signature is valid.
This is the case only for seesaw models in which $R_{ll}$ is close to one in this region, hence to the bottom left of the $R_{ll}$ bands of the \BM under consideration.
The reach of the $\HLLHC$ \cite{Drewes:2019fou}, \FCC-$ee$ \cite{Chrzaszcz:2020emg, Alimena:2022hfr}, \LHeC and \FCC-$eh$ \cite{Antusch:2019eiz} are shown as black lines.
When producing the figure, we have neglected potential effects from a nonzero damping factor $\lambda$ as well as from times shorter than the Jacob-Sachs threshold; for details, see \cref{sec:external wave packets,sec:MadGraph}.
\dummyacronym{LHeC}
} \label{fig:Rll bands}
\end{figure}

Although neither pure Dirac nor single Majorana particles are sufficient to describe the phenomenology of low-scale seesaw heavy neutrinos, almost all prior searches are performed using these two models.
Therefore it is necessary to correctly interpret these searches in order to extract the proper bounds on the pseudo-Dirac heavy neutrinos of low-scale seesaw models.
Customarily, the searches relying on displaced vertex signatures are reported for both pure Dirac and single Majorana \HNLs.
Since these searches usually do not rely on \LNV, the major difference affecting these searches is the factor of two that appears in the decay width and is described in detail in \cref{sec:HNL characterisation}.
Since the Dirac particle has the same decay width as the pseudo-Dirac particle, it is prudent to interpret the exclusion extracted for displaced Dirac \HNLs as the correct exclusions for displaced pseudo-Dirac heavy neutrinos.
The situation is more complicated in the case of prompt searches.
The majority of prompt searches rely on the very clean same-sign dilepton signature of \LNV decays.
Since these are not generated for Dirac particles, the searches are only reported for single Majorana particles.
Therefore, these searches assume an $R_{ll}$ of one, which depending on the mass splitting, might not be a good approximation of reality.
Hence the exclusion bound depends on the details of the model under consideration.
In practice, one has to check the intersection between the $R_{ll}$ band of one specific model and the reported exclusion bounds for the same-sign dilepton search in a representation such as the one presented in \cref{fig:Rll bands}.
For $R_{ll}$ close to one (to the bottom left of the bands) the reported exclusion bounds are valid up to the factor of two between the simulated single Majorana particle and the realistic pseudo-Dirac particle discussed in \cref{sec:HNL characterisation}.
For $R_{ll}$ close to zero (to the top right of the bands) there will be no \LNV, rendering the searches inconsequential.
Therefore, prompt searches can only be interpreted model-dependently, \eg as a function of $\Delta m$.
The model file presented in \cref{sec:FeynRules} together with the example implementation of \NNOs in \cref{sec:MadGraph} provides a unified framework for the correct simulation of prompt and displaced processes of pseudo-Dirac heavy neutrinos in low-scale seesaw models.

\section{Conclusion} \label{sec:conclusion}

In this paper, we have reviewed the arguments that collider testable seesaw models that do not rely on tuning must be protected by an approximate \LNLS.
The breaking of the symmetry by small parameters simultaneously ensures small \LNV, tiny \SM neutrino masses, and generically generates a small mass splitting between the heavy neutrinos.
The latter leads to \NNOs that can have macroscopic oscillation lengths such that they are potentially resolvable at high-energy colliders.

In order to simulate the effects of these oscillations, we have reviewed the \SPSS that builds on a systematic expansion in the small parameter governing the breaking of the \LNLS.
We have then reduced it to the \pSPSS that relies on the minimal number of parameters necessary to describe pseudo-Dirac heavy neutrinos and \NNOs in phenomenological studies.
In addition to the \SM parameters, the \pSPSS introduces only the Majorana mass parameter $m_M$, the active-sterile mixing parameters $\vec \theta = \row{\theta_e,\theta_\mu,\theta_\tau}^\trans$, the mass splitting $\Delta m$ governing the oscillation period, and a damping parameter $\lambda$ that captures possible decoherence effects.
The last of these can be neglected in the parts of the parameter space where prospects are good to resolve the oscillations, as discussed in \cite{Antusch:2020pnn} and further elaborated in \cite{Antusch:2023nqd}.

In order to facilitate \MC studies using this model, we have published a \software{FeynRules} implementation of the \pSPSS together with a patch extending \software{MadGraph} in such a way that it is capable of simulating \NNOs.
Using these tools, we have calculated some example results:
We have demonstrated that our implementation recovers typical prior results, such as the heavy neutrino decay width.
Additionally, we have provided the maximal mass splitting $\Delta m$ for which it may be feasible to resolve the oscillation pattern when only a fraction of events containing the largest Lorentz factors is considered.
Furthermore, we have shown that the integrated effect of the oscillations is reproduced by our model file.
We have emphasized in this context that care has to be taken when measuring $R_{ll}$, as the finite detector geometry can have a major impact on the derived values.
Finally, we have demonstrated the dependence of the impact parameter on the angular-dependent spin correlation and have shown that this observable can be used, in principle, to distinguish between \LNC and \LNV decays.

\subsection*{Acknowledgements}

The work of J.H.\ was partially supported by the Portuguese Fundação para a Ciência e a Tecnologia (FCT) through the projects CFTP-FCT unit UIDB/\allowbreak00777/\allowbreak2020 and UIDP/\allowbreak00777/\allowbreak2020.
S.A.\ and J.R.\ acknowledge partial support from the Swiss National Science Foundation grant 200020/175502.

\appendix

\section{Decay widths of Majorana, Dirac, and pseudo-Dirac particles} \label{sec:HNL characterisation}

In the main part of this article, we have discussed that in order to explain light neutrino masses, the corresponding heavy neutrino cannot be a Dirac particle.
Additionally, in order to generate the observed light neutrino oscillation pattern, at least two neutrinos have to be added to the \SM.
Nevertheless, it is standard practice to study the phenomenology of a single heavy neutrino that is either of Majorana or Dirac type.
In the following, we compare their phenomenology with the one of pseudo-Dirac heavy neutrinos with different mass splittings and comment on the decay widths of each of these particles.

\paragraph{Single Majorana}

The relevant part of the Lagrangian describing a single Majorana particle $N_1$ reads
\begin{equation}
\mathcal L_\text{Majorana} = - y_\alpha \widebar{N_1^c} \widetilde H^\dagger \ell_\alpha - \frac12 m_M^{} \widebar{N_1^c} N_1^{} + \HC\,.
\end{equation}
After \EWSB, the coupling to the $W$ bosons is obtained from the kinetic terms of the active neutrinos via the mixing of light $\nu_\alpha$ and heavy $n_4$ neutrino mass eigenstates and reads
\begin{align}\label{eq:coupling_majorana_W}
\mathcal L_\text{Majorana}^W &= \theta_\alpha'^* \widebar l_\alpha n_4 W + \HC \,, &
& \coupling{n_4}{l_\alpha}{\theta_\alpha^\prime}\,.
\end{align}
where the mixing parameter is given by $\vec \theta' = \vec y v m_M^{-1}$.
The expected number of \LNC and \LNV events at a collider experiment can then be schematically computed using
\begin{equation}
\abs{\mathcal A}^2 = \abs*{\scatterLNC{\theta_\alpha'}{m}}^2 + \abs*{\scatterLNV{\theta_\alpha'}{m}}^2 \propto 2 \abs{\theta_\alpha^\prime}^2 \,,
\end{equation}
where the diagram encodes the production and decay of a heavy neutrino from and to charged leptons and $W$ bosons.
In this case, one expects an equal number of \LNC and \LNV events and therefore $R_{ll}=1$.
As long as $\Gamma \ll m$, the magnitude of the (greyed out) coupling in the decay vertex is not relevant since it is cancelled by the decay width dependence of the propagator of the heavy neutrino, which can be verified using the narrow width approximation.

In general, the value of the active-sterile mixing parameter can be obtained by measuring the production cross section and the decay width of the heavy neutrino.
In the case of a single Majorana particle, the production cross section is proportional to $\abs{\vec \theta^\prime}^2$.
Generically, the decay width can be obtained by scanning the $p^2$ distribution of any process with the heavy neutrino mass eigenstate on the $s$-channel and comparing it to the Breit-Wigner shape.
Furthermore, regarding long-lived particles, it can be extracted from the decay exponential reconstructed by counting the number of displaced vertices per distance interval.
In the case of a single Majorana particle, the decay width is proportional to $\abs{\vec \theta^\prime}^2$.

\paragraph{Dirac}

A heavy Dirac neutrino can be described as two mass-degenerate Majorana neutrinos $N_{\nicefrac12}$ with an exact phase difference of $-i$ in their couplings.
The relevant part of the Lagrangian is given by
\begin{equation}
\mathcal L_\text{Dirac} = - y_\alpha \widebar{N_1^c} \ell_\alpha \widetilde H^\dagger - m_M^{} \widebar{N_1^c} N_2^{} + \HC\,.
\end{equation}
After \EWSB, the mass matrix of the neutrinos can be diagonalised, and the active \SM neutrino interaction eigenstates can be expressed as a linear combination of the heavy neutrino mass eigenstates.
These linear combinations are the heavy neutrino $N$ and heavy antineutrino $\widebar N$ introduced in \cref{sec:symmetric seesaw}
\begin{align}
N &= \frac{\theta^*}{\sqrt2} \left(-i n_4 + n_5 \right) \,, &
\widebar N &= \frac{\theta^*}{\sqrt2} \left(i n_4 + n_5 \right)\,,
\end{align}
which leads to the coupling to the gauge bosons via
\begin{align} \label{eq:Dirac mass eigenstate}
\mathcal L_\text{Dirac}^W &= \theta_\alpha^* \widebar l_\alpha \frac{1}{\sqrt2} \left(- i n_4 + n_5 \right) W  + \HC \,, &
& \coupling{n_4}{l_\alpha}{\pm i \frac{\theta_\alpha}{\sqrt2}} \,, &
\coupling{n_5}{l_\alpha}{\frac{\theta_\alpha}{\sqrt2}}\,,
\end{align}
where $\vec \theta = \vec y v m_M^{-1}$.
Compared to the case of a single Majorana particle, the coupling of the gauge bosons to each heavy neutrino mass eigenstate obtains an additional factor of $\flatfrac{1}{\sqrt2}$, which originates from the diagonalisation of the heavy neutrino $2\times2$ block in the mass matrix.
The number of \LNC and \LNV events can be derived using the Feynman diagrams
\begin{equation}\label{eq:feynman_dirac}
\begin{aligned}
\abs{\mathcal A}^2 &=
\abs*{\scatterLNC[\frac{(\pm i\theta_\alpha)^*}{\sqrt 2}]{\frac{\pm i\theta_\alpha}{\sqrt 2}}{m_4} + \scatterLNC[\frac{\theta_\alpha^*}{\sqrt 2}]{\frac{\theta_\alpha}{\sqrt 2}}{m_5}}^2 +
\abs*{\scatterLNV{\frac{\pm i\theta_\alpha}{\sqrt 2}}{m_4} + \scatterLNV{\frac{\theta_\alpha}{\sqrt 2}}{m_5}}^2 \\
& = \abs*{\scatterLNC[\frac{(\pm i\theta_\alpha)^*}{\sqrt 2}]{\pm i\frac{\theta_\alpha}{\sqrt 2}}{m_4} + \scatterLNC[\frac{\theta_\alpha^*}{\sqrt 2}]{\frac{\theta_\alpha}{\sqrt 2}}{m_5}}^2 = 4 \abs*{\scatterLNC[\frac{\theta_\alpha^*}{\sqrt 2}]{\frac{\theta_\alpha}{\sqrt 2}}{m_{\nicefrac45}}}^2 \propto 4 \abs*{\frac{\theta_\alpha}{\sqrt2}}^2 = 2 \abs{\theta_\alpha}^2 \,.
\end{aligned}
\end{equation}
where the Dirac condition ensures mass degeneracy $m_4 = m_5$.
In contrast to the Majorana heavy neutrino, the Dirac heavy neutrino generates only \LNC decays and, therefore, $R_{ll} = 0$.
In the picture where the heavy Dirac neutrino is represented by two Majorana \DOFs, this is due to the phase difference of $-i$ between the couplings of the mass eigenstates, ensuring that the \LNV diagrams cancel exactly.
Both the Dirac and the Majorana heavy neutrino yield the same total number of events, such that it is not possible to distinguish them just by the observation of the overall number of events.

The production cross section of a single mass eigenstate $n_{\nicefrac45}$ in \eqref{eq:Dirac mass eigenstate} is proportional to $\abs{\vec \theta}^2/2$.
However, since it is intrinsically impossible to distinguish the two identical mass eigenstates in a single Dirac particle, the measured production cross section is proportional to $\abs{\vec \theta}^2$.
Nonetheless, the decay width is only proportional to $\abs{\vec \theta}^2/2$ as one does not sum over the two \DOFs in order to calculate the total decay width.
Therefore, in comparison to the single Majorana heavy neutrino, the decay width is reduced by a factor of two.

Naively one might expect to distinguish Dirac from Majorana heavy neutrinos in this way.
However, this is impossible as we will demonstrate using different limits of a model consisting of a pair of two heavy Majorana neutrinos with finite mass splitting.

\paragraph{Majorana pair}

The Majorana and Dirac heavy neutrinos can both be described as limiting cases of a model with two Majorana particles also encompassing the pseudo-Dirac heavy neutrino.
Starting from the Dirac heavy neutrino, a small perturbation of the \LNLS, such that the two masses of the Majorana \DOFs, as well as the modulus of their couplings, are no longer exactly degenerate, generates a pseudo-Dirac heavy neutrino, as explicitly shown in \cref{sec:small breaking}.
Conversely, the Dirac heavy neutrino can be seen as the limit of the pseudo-Dirac heavy neutrino in which the perturbation of the \LNLS goes to zero, such that the symmetry is restored.
In the limit of large mass splitting, the phenomenology of two well-separated Majorana heavy neutrinos is recovered.
In the following, when we refer to the size of the mass splitting, we imply that the Yukawa couplings are also perturbed compared to the degeneracy given in the Dirac case.

\subparagraph{Pseudo-Dirac}

For small mass splittings as described in \cref{sec:small breaking}, the mass eigenstates $n_{\nicefrac45}$ in \eqref{eq:feynman_dirac} are still produced coherently describing the phenomenology of a pseudo-Dirac pair.
However, the \LNV diagrams do not cancel exactly anymore, leading to \LNV processes.
While the total number of produced events stays the same, \NNOs are possible such that a part of generated events are now \LNV leading to $0< R_{ll} < 1$.
However, the arguments for the production and decay of a pure Dirac particle apply also to this case.

\subparagraph{Intermediate mass splitting}

\begin{table}
\begin{tabular}{l*5c} \toprule
 & Pure Dirac & \multicolumn{3}{c}{Majorana Pair} & Single Majorana \\ \cmidrule{3-5}
 & & Pseudo-Dirac & & \\ \cmidrule{3-3}
Mass splitting & & Small & Intermediate & Large \\ \midrule
Production & $\abs{\vec \theta}^2\hphantom{/2}$ & $\abs{\vec \theta}^2\hphantom{/2}$ & $\abs{\vec \theta}^2\hphantom{/2}$ & $\abs{\vec \theta^\prime}^2$ & $\abs{\vec \theta^\prime}^2$ \\
Decay & $\abs{\vec \theta}^2/2$ & $\abs{\vec \theta}^2/2$ & $\abs{\vec \theta}^2/2$ & $\abs{\vec \theta^\prime}^2$ & $\abs{\vec \theta^\prime}^2$ \\
$R_{ll}$ & $0$ & $0<R_{ll}<1$ & $1$ & $1$ & $1$ \\
\bottomrule \end{tabular}
\caption[Production ratios and decay widths for different types of heavy neutrinos]{
Production ratios and decay widths for different types of heavy neutrinos.
The primed mixing parameters occur in models without mass mixing between two heavy states, while the unprimed mixing parameters occur in models with mass mixing between the two heavy neutrinos.
} \label{tab:neutrino types}
\end{table}

While intermediate mass splitting can still be too small to experimentally distinguish different mass eigenstates, it is still possible that decoherence already sets in.
This effect is governed by the observability conditions \cite{Antusch:2020pnn,Antusch:2023nqd} and by the damping factor included in the patch.
In this case, the interference between diagrams with different mass eigenstates in \eqref{eq:Dirac mass eigenstate} is destroyed.
Therefore the \LNC and \LNV processes become both equally likely while the total number of events is proportional to
\begin{multline}
\abs{\mathcal A}^2 =
\abs*{\scatterLNC[\frac{(\pm i\theta_\alpha)^*}{\sqrt 2}]{\frac{\pm i\theta_\alpha}{\sqrt 2}}{m_4}}^2 + \abs*{\scatterLNC[\frac{\theta_\alpha}{\sqrt 2}]{\frac{\theta_\alpha}{\sqrt 2}}{m_5}}^2 +
\abs*{\scatterLNV{\frac{\pm i\theta_\alpha}{\sqrt 2}}{m_4}}^2 + \abs*{\scatterLNV{\frac{\theta_\alpha}{\sqrt 2}}{m_5}}^2 \\
 \propto 4 \abs*{\frac{\theta_\alpha}{\sqrt2}}^2 = 2 \abs{\theta_\alpha}^2
\end{multline}
and thus remains the same.
The production cross section corresponds to the sum of the production cross sections of both mass eigenstates, \ie it is proportional to $\abs{\vec\theta}^2$, while the decay width is the one of each individual mass eigenstate and therefore proportional to $\abs{\vec \theta}^2/2$.
Although the ratio between \LNC and \LNV processes is identical to the case of a single heavy Majorana neutrino, one still obtains a decay width that is a factor of two smaller compared to the case of a single Majorana particle.
Therefore, from the comparison of the decay width and the production cross section of a newly discovered heavy neutrino, one can gain information about the number of Majorana \DOFs taking part in the production cross section.
However, it is not possible to distinguish a Dirac particle from two Majorana particles with a small or intermediate mass splitting
just by the comparison of the production cross sections, decay widths or total event numbers.

\subparagraph{Large mass splitting}

For large mass splitting, the individual mass eigenstates can be experimentally distinguished.
As before, one finds for each mass eigenstate the production cross section as well as the decay width to be approximately proportional to $\abs{\vec \theta}^2/2$.
Considering that it is likely that the lighter of the two mass eigenstates is discovered first, it is insightful to express the coupling in the language of a single Majorana, which would lead to $\abs{\vec \theta^\prime}^2 =: \abs{\vec \theta}^2/2$.
Therefore, one obtains exactly the same phenomenology as if the two mass eigenstates are two distinct Majorana particles, avoiding the unnecessary interpretation as a quasi pseudo-Dirac particle with large mass splitting.

\paragraph{Summary}

We summarise the discussion above in \cref{tab:neutrino types} and conclude by noting that the comparison of the decay widths yields insight into whether there are two Majorana \DOFs, as in the case of Dirac particles, or only one, as in the case of a single Majorana particle, but it cannot prove the Dirac property of the heavy neutrino.
This is already clear from the fact that the factor of two in the decay widths approximately remains when a small \LNV term is added, which turns the pure Dirac heavy neutrino into a pseudo-Dirac heavy neutrino.
In order to test the Dirac property of the heavy neutrino, it is unavoidable to test for \LNV.
From a measurement of the mass(es) and lifetime(s), one can, at best, conclude that there are two Majorana particles with almost identical properties.
But whether these two \DOFs can be combined to form a \LNC Dirac heavy neutrino depends on the phase between the Yukawa couplings of the two mass-degenerate Majorana particles.
Only when the phase is $-i$, then, on the one hand, \LNV cancels out and on the other hand, the two Majorana particle \DOFs can be combined into a Dirac heavy neutrino.
As long as this phase (related to \LNV) is not tested, one can only speak of two Majorana \DOFs.

\begin{lstlisting}[
float,captionpos=b,caption={[Original \software{MadGraph} code]%
Original \software{MadGraph} code calculating particle \TOF.%
},label=lst:original
]
for event in lhe:
        for particle in event:
            id = particle.pid
            width = param_card['decay'].get((abs(id),)).value
            if width:
                vtim = c * random.expovariate(width/cst)
                if vtim > threshold:
                    particle.vtim = vtim
        #write this modify event
        output.write(str(event))
    output.write('</LesHouchesEvents>\n')
    output.close()
\end{lstlisting}

\begin{lstlisting}[
 otherkeywords={8000011,8000012},
 emph={8000011,8000012},
 emphstyle={\itshape\bfseries},
 float,
 captionpos=b,
 caption={[Patched \software{MadGraph} code]%
Patched \software{MadGraph} code calculating particle \TOF and \NNOs.
Note that the sterile neutrino particle IDs and the position of the \texttt{mass splitting} parameter in the \nolinkurl{param.card} are model file dependent and might need to be adjusted.
%
}, label=lst:patch
]
mass_splitting = param_card.get_value('PSPSS', 2)|\label{ln:block}|
damping = param_card.get_value('PSPSS', 6)|\label{ln:block_damping}|
for event in lhe:
    leptonnumber = 0
    write_event = True
    for particle in event:|\label{ln:particle loop}|
        if particle.status == 1:
            if particle.pid in [11, 13, 15]:
                leptonnumber += 1
            elif particle.pid in [-11, -13, -15]:
                leptonnumber -= 1
    for particle in event:
        id = particle.pid
        width = param_card['decay'].get((abs(id),)).value
        if width:
            if id in [8000011, 8000012]:|\label{ln:ids}|
                tau0 = random.expovariate(width / cst)|\label{ln:tau0}|
                if 0.5 * (1 + math.exp(-damping)*math.cos(mass_splitting * tau0 / cst)) >= random.random():|\label{ln:LNC}|
                    write_event = (leptonnumber == 0)
                else:
                    write_event = (leptonnumber != 0)
                vtim = tau0 * c|\label{ln:vtim}|
            else:
                vtim = c * random.expovariate(width / cst)
            if vtim > threshold:
                particle.vtim = vtim
    # write this modify event
    if write_event:|\label{ln:write event}|
        output.write(str(event))
output.write('</LesHouchesEvents>\n')
output.close()
\end{lstlisting}

\section{\software{MadGraph} Patch} \label{sec:patch}

The patch presented here is used to include \NNOs in \software{MadGraph}.
It is applied to the \nolinkurl{[pSPSS]/bin/internal/madevent_interface.py}
\footnote{
In other versions, the relevant section might be found in the \nolinkurl{[pSPSS]/bin/internal/common_run_interface.py} file.
}
generated by \software[2.9.10 (LTS)]{MadGraph5\_aMC@NLO}.
To apply the patch, the main event loop, responsible for adding \TOF information to the \nolinkurl{.lhe} file given in \cref{lst:original}, has to be replaced by the code presented in \cref{lst:patch}.
\footnote{
The relevant lines can be found by searching for the keyword \texttm{vtim}.
}
The loop over the particles in the event, starting in \cref{ln:particle loop}, is used to determine the \code{leptonnumber} of the event.
Electrons, muons and $\tau$-leptons are counted as leptons, whereas their antiparticles are counted as antileptons.
The \code{masssplitting} is extracted from the \code{PSPSS} block in the \nolinkurl{param.card} in \cref{ln:block} and the damping factor \code{damping} similar in \cref{ln:block}.
In \cref{ln:ids} the program executes the new code of the patch if the particle is a heavy neutrino with particle ID \code{8000011} or \code{8000012} as defined in the model file presented in \cref{sec:FeynRules}.
First, the proper time at which the heavy neutrino decays is obtained in \cref{ln:tau0}.
Next, the formula for the oscillation probability in the proper time frame is used together with a random number to decide if the event should be \LNC or \LNV.
Subsequently, if the \code{leptonnumber} agrees with the result from \cref{ln:LNC}, the switch \code{write_event} is set to \code{True}.
Otherwise, it is set to \code{False}.
In \cref{ln:vtim}, the proper time is converted to \unit{mm} as in the unpatched code from \software{MadGraph}.
Finally, using the \code{write_event} switch from above, it is decided if the event should be kept or discarded in \cref{ln:write event}.

Note that when using the automated width computation of \software{MadGraph}, for some parameter points, there are issues with the computation of the phase space volume in the decay $n_5 \rightarrow n_4 \nu \nu$, where $n_5$ is the heaviest neutrino, and the $\nu$ are massless neutrinos, due to the unusually small mass splitting.
Since we found this decay channel to be negligible, we fixed the issue by replacing the argument of the square root of the return value of the function \code{calculate_apx_psarea} in the file \nolinkurl{[MadGraph]/mg5decay/decay_objects.py} by its absolute value.
However, this can be a drastic change in the \software{MadGraph} behaviour.
Therefore we have checked that the results are physically meaningful \cite{launchpad}.
In our case, that was found to be true since the problematic decay channel is discarded by \software{MadGraph} at a later step independently of the change, so there is no impact on any physical results.
Additionally, we also compared the automated computed width for a parameter point with a small mass splitting of the heavy neutrinos to a parameter point with a large mass splitting, changing the other parameters in such a way that the mass of $n_4$, as well as its couplings constant, stays constant.
The two computations yield the same results, showing that there is no impact from the extremely small mass splitting in the computation performed by \software{MadGraph}.

\printbibliography

\end{document}